\documentclass{aa}
\usepackage[dvips]{graphicx}
\usepackage{txfonts}
\usepackage{bm} 

\begin{document}
%
%
\newcommand{\nc}{\newcommand}
\nc{\bea}{\begin{eqnarray}} \nc{\eea}{\end{eqnarray}}
\nc{\beq}{\begin{equation}} \nc{\eeq}{\end{equation}}
\nc{\ve}[1]{{\mathbf{#1}}}
 \nc{\beps}{\boldsymbol{\varepsilon}}
 \nc{\bP}{\mathbf{P}}
 \nc{\bF}{\mathbf{F}}
 \nc{\bB}{\mathbf{B}}
 \nc{\bZ}{\mathbf{Z}}
 \nc{\bI}{\mathbf{I}}
 \nc{\bIt}{\mathbf{I}_t}
 \nc{\bIm}{\mathbf{I}_m}
 \nc{\bA}{\mathbf{A}}
 \nc{\bC}{\mathbf{C}}
 \nc{\bN}{\mathbf{N}}
 \nc{\bH}{\mathbf{H}}
 \nc{\binvN}{\mathbf{N}^{-1}}
 \nc{\fk}{f_\mathrm{k}}
 \nc{\etal}{{et al.}}


\title{Comparison of map-making algorithms for CMB experiments}

\author{T. Poutanen\inst{1,2}, G. de Gasperis\inst{3}, E. Hivon\inst{4},
H. Kurki-Suonio\inst{2}, A. Balbi\inst{3,5}, J. Borrill\inst{6,7},
C. Cantalupo\inst{7,6}, O. Dor\'e\inst{8}, E. Keih\"anen\inst{1,2},
C. Lawrence\inst{9}, D. Maino\inst{10}, P. Natoli\inst{3,5}, S.
Prunet\inst{11}, R. Stompor\inst{6,7}, and R. Teyssier\inst{12}}

\offprints{T.~Poutanen, \email{torsti.poutanen@helsinki.fi}}

\institute{Helsinki Institute of Physics, P.O. Box 64, FIN-00014
Helsinki, Finland \and University of Helsinki, Department of
Physical Sciences, P.O. Box 64, FIN-00014, Helsinki, Finland \and
Dipartimento di Fisica, Universit\`{a} di Roma ``Tor Vergata'', via
della Ricerca Scientifica 1, I-00133 Roma, Italy \and IPAC, MS
100-22, Caltech, Pasadena, CA 91125, U.S.A  \and INFN, Sezione di
Roma II, via della Ricerca Scientifica 1, I-00133 Roma, Italy \and
Computational Research Division, Lawrence Berkeley National
Laboratory, Berkeley CA 94720, U.S.A. \and Space Sciences
Laboratory, University of California Berkeley, Berkeley CA 94720,
U.S.A. \and Department of Astrophysical Sciences, Princeton
University, Princeton, NJ 08540, U.S.A. \and Jet Propulsion
Laboratory, 4800 Oak Grove Drive, Mailstop 169-327, Pasadena CA
91109, U.S.A. \and Dipartimento di Fisica, Universit\`{a} di Milano,
Via Celoria 16, I-20131 Milano, Italy \and Institut d'Astrophysique
de Paris, 98 bis Boulevard Arago, F-75104 Paris, France \and Service
d'Astrophysique, DAPNIA, Centre d'Etudes de Saclay, F-91191
Gif-sur-Yvette, France}

\date{September 20, 2005}

\abstract{We have compared the cosmic microwave background (CMB)
temperature anisotropy maps made from one-year time ordered data
(TOD) streams that simulated observations of the originally planned
100~GHz {\sc Planck} Low Frequency Instrument (LFI). The maps were
made with three different codes.  Two of these, ROMA and MapCUMBA,
were implementations of maximum-likelihood (ML) map-making, whereas
the third was an implementation of the destriping algorithm. The
purpose of this paper is to compare these two methods, ML and
destriping, in terms of the maps they produce and the angular power
spectrum estimates derived from these maps. The difference in the
maps produced by the two ML codes was found to be negligible.  As
expected, ML was found to produce maps with lower residual noise than
destriping.  In addition to residual noise, the maps also contain an
error which is due to the effect of subpixel structure in the signal
on the map-making method. This error is larger for ML than for
destriping. If this error is not corrected a bias will be introduced
in the power spectrum estimates. This study is related to {\sc
Planck} activities.

}

\authorrunning{Poutanen et al.}

\maketitle


\section{Introduction}

Map-making from an observed time ordered data (TOD) stream is an
important step in the data processing pipeline of a cosmic microwave
background (CMB) experiment. A number of map-making algorithms
which, under the assumption of Gaussian distributed and stationary
noise, aim at finding the optimal minimum variance map have been
proposed (Wright~\cite{Wri96}; Borrill~\cite{Bor99}; Dor\'e et
al.~\cite{Dor01}; Natoli et al.~\cite{Nat01}). The destriping
technique (Burigana et al.~\cite{Bur97}; Delabrouille~\cite{Del98};
Maino et al.~\cite{Mai99},~\cite{Mai02}; Revenu et
al.~\cite{Rev00a},~\cite{Rev00b}; Keih\"anen et al.~\cite{Kei04}) is
a simpler map-making method.

For this study we utilized simulated one-year TOD streams, that had
been produced in the course of the work of the CTP ($C_\ell$ for
Temperature and Polarisation) Working Group of the {\sc Planck}
Consortia. These TODs represent the output of a single detector of
the originally planned 100 GHz LFI (Low Frequency Instrument) channel
of the {\sc Planck} satellite. This simulated data contains
contributions from CMB and foreground emissions, as well as from the
instrumental noise. The noise was assumed Gaussian distributed and
stationary over the entire mission, containing a $1/f$ and a white
noise component. We considered temperature anisotropies only; no
polarisation.

Output maps were generated from the TODs using three distinct
map-making codes. The output maps and their angular power spectra
were critically compared. The map-making codes were ROMA (Roma
Optimal Mapmaking Algorithm; Natoli et al.~\cite{Nat01}; de Gasperis
et al.~\cite{deG05}), MapCUMBA (originally introduced by Dor\'e et
al.~\cite{Dor01}, current version based on the preconditioned
conjugate gradient method) and destriping (Keih\"anen et
al.~\cite{Kei04}). All methods have been developed to treat {\sc
Planck}-like data.

A minimum variance map maximizes the likelihood function involving
the full noise covariance. The ROMA and MapCUMBA algorithms aim at
producing the minimum variance map. To accomplish this, the
algorithms require knowledge on the characteristics of the instrument
noise. Both iterative (Dor\'e et al.~\cite{Dor01}) and non-iterative
(Natoli et al.~\cite{Nat02}) methods to estimate the noise properties
directly from the data have been proposed. In this paper ROMA and
MapCUMBA are referred to with a common name maximum likelihood (ML)
map-making.

Destriping does not employ the noise covariance matrix and does not
aim at producing a minimum variance map in this sense.  Thus it does
not require prior knowledge on the characteristics of the instrument
noise. This simplifies the algorithm considerably as compared to the
ML map-making. In spite of this, destriping is able to provide an
estimate of the low-frequency part of the instrument noise and to
return a TOD where these noise components have been removed.
Destriping can also be applied to estimate various systematic effects
and drifts, remove them and return a cleaned TOD (see e.g. Mennella
et al.~\cite{Men02}).

The aim of this paper is to compare these two methods, ML (ROMA and
MapCUMBA) and destriping, in terms of the maps they produce and the
angular power spectrum ($C_\ell$) estimates derived from these maps.

This paper is organized as follows. In Sect.~\ref{sec:method} we
describe the map-making and power spectrum estimation methods applied
in this study. The simulated one-year TOD streams are introduced in
Sect.~\ref{sec:tod}. The output maps are compared in
Sect.~\ref{sec:maps} and the power spectrum estimates produced from
the simulated TODs are examined in Sect.~\ref{sec:estimates}. The
conclusions are given in Sect.~\ref{sec:conclu}. In
Appendix~\ref{sec:recerror} we describe how the output map is split
in the (wanted) binned noiseless map and in the (unwanted)
reconstruction error map. These quantities were considered in the
comparison of the output maps in Sect.~\ref{sec:maps}.  In Appendix B
we discuss some details about how the $C_\ell$ of the maps are
related to the input $C_\ell$ used to generate the TODs.

\section{Methods}
\label{sec:method}

Let us denote by a column vector $\ve{y}$ the samples of the observed
TOD. The length of $\ve{y}$ is $N_t$, the number of samples in the
total mission. In the map-making problem we assume that the signal
samples are scanned from a pixelized temperature map ($\ve{m}$).
(This assumption of course represents an approximation to reality,
and thus contributes to error in the final maps.) The length of the
column vector $\ve{m}$ is $N_{\rm pix}$, the number of pixels in the
map.

The scanning is implemented by a pointing matrix $\bP$. The size of
the pointing matrix is $[P]=(N_t,N_{\rm pix})$. Each row contains
zeros except for one element with value one, indicating the pixel at
which the detector beam centre was pointing when the sample was
measured.  The map-making methods discussed here utilize pointing
information only to the accuracy given by the pixel size. If the
instrument beam response is spherically symmetric (and as we are
neglecting polarisation), information about the rotation angle of
the detector around the line of sight is not needed, and such a
simple pointing matrix is sufficient.  The pixel temperature of the
output map will then represent a convolution of the sky and the
beam. We have used in this study simulated TODs corresponding to
both spherically symmetric and elliptic beams; but a study of the
effects of beam shape is postponed to a future detailed study by the
CTP group.

For an asymmetric beam every pixel will be smoothed with a different
response that is the mean of the beam orientations of the
observations falling in that pixel. Recently a deconvolution
map-making algorithm was introduced that can provide an output map
where the smoothing of the instrument beam (symmetric or asymmetric)
is deconvolved leading to a map that is an estimate of the true sky
(Armitage \& Wandelt~\cite{Arm04}). Deconvolution map-making is
beyond the scope of this study.

\subsection{ROMA and MapCUMBA}
\label{subsec:igls}

The output map ($\ve{m}$) is solved by minimizing the
log-likelihood formula (Natoli et al.~\cite{Nat01})
 \beq
   \chi^2 = (\ve{y} - \bP\ve{m})^T\binvN(\ve{y} - \bP\ve{m}). \label{igls1}
 \eeq
Here $\bN$ is the noise covariance matrix $\bN =
\langle\ve{n}\ve{n}^T\rangle$, where $\ve{n}$ is the instrument noise
component of the TOD and $\langle\cdot\rangle$ denotes expectation
value. A set of linear equations is obtained for the output map
 \beq
 \bP^T\binvN\bP\ve{m} = \bP^T\binvN\ve{y}. \label{igls2}
 \eeq
Eq. (\ref{igls2}) is a general result for the minimum variance map.

Usually ML map-making assumes the noise to be stationary throughout
the mission. It is further assumed that the elements of the
covariance matrix ($N_{ij}$) vanish when $|i-j|$ is larger than some
$N_{\eta}$ and $N_{\eta} \ll N_t$. This means that the correlation
is significant only across a number of samples that is a tiny
fraction of the total length of the TOD. Thus the noise correlation
matrix $\bN$ can be approximated by a circulant matrix (Natoli et
al.~\cite{Nat01}). Note that the matrix $\binvN$ is approximately
circulant as well. The multiplication $\binvN\ve{y}$ can be carried
out more easily in the frequency domain where $\binvN$ is diagonal
(Natoli et al.~\cite{Nat01}). Both in the ROMA and in the MapCUMBA
algorithms the output map is solved from Eq. (\ref{igls2}) with an
iterative preconditioned conjugate gradient method. The iterations
are repeated until the fractional difference has reached a low
enough value (Natoli et al.~\cite{Nat01}). This limit is typically
on the order of $10^{-6}$.

Due to the circulant matrix approximation each row of the matrix
$\binvN$ contains the same element values with a different cyclic
permutation. It is assumed that only the elements of a row with
$|i-j| \le N_{\xi}$ ($N_{\xi} \ll N_t$) have non-zero values. The
rest of the elements are zero. The collection of the non-zero
elements (of a row) is called the {\it noise filter}. The lag of an
element is the difference of its indices ($i-j$). The choice of the
value $N_{\xi}$ is a significant decision for the quality of the
output maps and for the computation time of the algorithm (Natoli et
al.~\cite{Nat01}).

Natoli et al. (\cite{Nat01}) have shown how ML map-making can be
applied when the instrument noise is only piece-wise stationary.

\subsection{Destriping technique} \label{subsec:destr}

The destriping technique for map-making has been derived from the
COBRAS/SAMBA Phase-A study (Bersanelli et al.~\cite{Ber96}). It has
been implemented by several groups (Burigana et al.~\cite{Bur97};
Delabrouille~\cite{Del98}; Maino et al.~\cite{Mai99},~\cite{Mai02};
Revenu et al.~\cite{Rev00a},~\cite{Rev00b}; Keih\"anen et
al.~\cite{Kei04}; Efstathiou~\cite{Efs05}). The implementation
studied in this paper (Keih\"anen et al.~\cite{Kei04}) makes use of
the fact that {\sc Planck} is a spinning spacecraft. Detector beams
are drawing almost great circles on the sky. Each scan circle is
observed several times before the spin axis is repointed. In order
to reduce the level of instrumental noise, the signal can be
averaged (``coadded'') over these scan circles.  In the following we
call this averaged scan circle a {\em ring}.  This coadding shortens
the TOD by a factor given by the number of circles between
repointings.

Janssen et al. (\cite{Jan96}) has pointed out that the effect of the
instrumental noise, in particular $1/f$ noise, on the ring can be
approximated by a uniform offset or ``baseline''. The key problem in
destriping is to find the amplitudes of these baselines. The
destriping technique uses the redundancy of the observing strategy by
considering the intersections (crossing points) between the scan
circles to obtain these amplitudes. A crossing point is defined as
two samples from different rings falling on the same pixel. According
to this definition, two closely spaced rings may have a crossing
point (or a sequence of them, when the rings run parallel) without
actually crossing each other, if the distance between the rings is
smaller than the pixel size.

The correlated noise component of the TOD is modelled as
$\ve{n}_{\rm{corr}} = \bF\ve{a}$ (Keih\"anen et al.~\cite{Kei04}).
Here vector $\ve{a}$ contains the amplitudes of the baselines and
matrix $\bF$ unfolds them into a TOD.  That is, $\bF\ve{a}$ is a
piecewise constant TOD, where the constants are given by the elements
of $\ve{a}$.  Once the amplitudes have been solved, $\bF\ve{a}$ is
subtracted from the original TOD to produce a cleaned TOD. Finally
the output map is binned from the cleaned TOD. This procedure is
formally described by minimizing the following likelihood function
with respect to the output map $\ve{m}$ and the amplitudes $\ve{a}$
(Keih\"anen et al.~\cite{Kei04})
 \beq
   \chi^2 = (\ve{y} - \bP\ve{m} - \bF\ve{a})^T(\ve{y}
   - \bP\ve{m} - \bF\ve{a})/\sigma^2. \label{destr1}
 \eeq
It is assumed here that the variance ($\sigma^2$) of the
non-correlated component of the noise is constant throughout the TOD.
The amplitudes ($\ve{a}$) can be solved from the equation
 \beq
   \bF^T\bZ\bF\ve{a} = \bF^T\bZ\ve{y} \label{destr2}
 \eeq
and the output map is given by
 \beq
   \ve{m} = (\bP^T\bP)^{-1}\bP^T(\ve{y} - \bF\ve{a}). \label{destr3}
 \eeq
In Eq. (\ref{destr2}) $\bZ \equiv \bIt-\bP(\bP^T\bP)^{-1} \bP^T$,
where $\bIt$ is a unit matrix with dimension $N_t$. The matrix $\bZ$
is determined by the crossing points of the scan circles. When $\bZ$
is acting on the TOD it subtracts from each sample the average of the
samples hitting the same pixel.

It turns out that the matrix $\bF^T\bZ\bF$ is singular and
additional conditions are required before the amplitudes can be
solved from Eq. (\ref{destr2}).  To accomplish this, we require that
the sum of the baseline amplitudes is zero.

The destriping method, as given by Eqs.~(\ref{destr2}) and
(\ref{destr3}) requires no knowledge of the noise power spectrum. (If
it is known that $\sigma^2$ varies between different parts of the
TOD, this information can of course be incorporated to improve the
accuracy of the method.)

The dimension of the matrix $\bF^T\bZ\bF$ equals the number of the
fitted baselines. A typical number of baselines for a one year TOD
(number of rings) is of the order of several thousands. This is a
less complex problem than solving the map in the ML method (cf. Eq.
(\ref{igls2})), because the number of pixels in the output map is
typically between $10^6 \ldots 10^8$.

Generalized approaches to destriping method have been implemented
(but not used in the present study) which are able to fit different
sets of base functions (in addition to the uniform baseline) and may
better remove the contributions of different systematic effects from
the TODs (Delabrouille~\cite{Del98}; Maino et al.~\cite{Mai02};
Keih\"anen et al.~\cite{Kei04},~\cite{Kei05}).

\subsection{Estimation of the angular power spectrum}
\label{subsec:powerspectrum}

We studied the CMB angular power spectrum estimates obtained from the
output maps. For the power spectrum estimation we used the MASTER
approach (Monte carlo Apodised Spherical Transform EstimatoR)
described by Hivon et al. (\cite{Hiv02}). MASTER has been adapted to
operate both with ROMA (Balbi et al.~\cite{Bal02}) and with
destriping (Poutanen et al.~\cite{Pou04}).

The angular power spectrum of the CMB sky $C_\ell^{\rm{in}}$ is
derived from the spherical harmonic expansion coefficients ($a_{\ell
m}$) of the sky
 \beq C_\ell^{\rm{in}} = \frac{1}{2\ell +1}
 \sum_{m=-\ell}^{\ell}{ |a_{\ell m}|^2}. \label{clin}
 \eeq
The simulated TODs are generated from such a spectrum, supposed to
represent the ``real sky''.  We call it the {\em input spectrum} for
our simulation + map-making problem. The $a_{\ell m}$ coefficients
are a realisation of the underlying ``theoretical'' angular power
spectrum $C_\ell^{\rm{th}}$. The ensemble mean (of many
realisations) equals the theoretical spectrum, $\langle
C_{\ell}^{\rm{in}}\rangle = C_\ell^{\rm{th}}$. The angular power
spectrum (pseudo spectrum) of an output map is denoted
$\widetilde{C}_\ell$. We further define $C_{\ell}^{\rm B}$, which is
the angular power spectrum of the binned noiseless map. That map is
obtained by binning the samples of the noiseless signal-only TOD to
map pixels.

The ensemble mean of the CMB pseudo spectrum depends on the ensemble
mean of the spectrum of the sky.  Hivon et al.~\cite{Hiv02} express
this relation as
 \beq
   \langle\widetilde{C}_\ell\rangle = \sum_{\ell'}M_{\ell
   \ell'} F_{\ell'} B^2_{\ell'}D^2_{\ell'} \langle C_{\ell'}^{\rm{in}}\rangle + \langle
   \widetilde{N}_\ell \rangle. \label{aps1}
 \eeq
The matrix $M_{\ell \ell'}$ is the mode coupling matrix (kernel
matrix) determined by the applied sky cut (Hivon et
al.~\cite{Hiv02}). A symmetric instrument beam is assumed and its
smoothing is modelled by $B_\ell^2$. Pixelisation introduces
additional smoothing, which is represented by the pixel window
factor $D_\ell^2$. The filter function $F_\ell$ represents a
possible distorting effect of map-making and the noise bias $\langle
\widetilde{N}_\ell \rangle$ the remaining noise.

There are a number of complications in the relation between the
spectrum of the output map $\widetilde{C}_\ell$ and that of the sky
$C_{\ell'}^{\rm{in}}$, not fully captured by Eq.~(\ref{aps1}). These
are related to the experimental setup, and, in the case of simulated
data, to imperfections in how the simulation models the experiment.
Some of these are discussed in Appendix \ref{sec:inputmap}. The
purpose of this study is to compare map-making algorithms. We want
to isolate the map-making errors from these other effects. Thus we
will not try to estimate the angular power spectrum of the sky
($C_\ell^{\rm{in}}$) but instead we compare the spectra of the
output maps to the spectrum of the binned noiseless map
($C_\ell^{\rm{B}}$). The map-making methods compared here are
derived from assumptions (see beginning of Sect.~\ref{sec:method})
which correspond to the binned noiseless map being equal to the true
sky (or the covered part of it), and thus it is effectively the
object the map-making methods are trying to estimate.

The spectra of the output map and the binned noiseless map can be
related by
 \beq
   \langle\widetilde{C}_\ell\rangle = F_{\ell} \langle{C}_{\ell}^{\rm B}\rangle
    + \langle \widetilde{N}_\ell \rangle. \label{aps2a}
 \eeq
Here $F_\ell$ accounts for the map-making errors only, and we shall
use this definition for the filter function. It can be determined by
e.g. signal-only MC simulations. We expect that its deviation from
one should be small. We have dropped the mode coupling matrix
because the sky coverage of the output map and the binned noiseless
map are identical.  Likewise, the beam (symmetric or not) and pixel
window have the same effect on both maps, and therefore do not
appear in Eq.~(\ref{aps2a}). The estimate of the spectrum of the
binned noiseless map ($\widehat{C}_{\ell}^{\rm B}$) can be obtained
by inverting the equation
 \beq
   \widetilde{C}_\ell = F_{\ell} \widehat{C}_{\ell}^{\rm B}
    + \langle \widetilde{N}_\ell \rangle \,. \label{aps2}
 \eeq

The influence of the instrument noise is modelled by the noise bias
term $\langle \widetilde{N}_\ell \rangle$. For destriping an analytic
method has been proposed (Efstathiou~\cite{Efs05}) that can provide
an estimate for the noise bias. However, in this study we used Monte
Carlo (MC) simulations to obtain an estimate for it. A number of
noise only TODs were generated from the power spectral density (PSD)
of the instrument noise. Maps were made from these TODs and a mean of
their pseudo spectra was derived. This mean is an estimate of the
noise bias.

\section{Time ordered data}
 \label{sec:tod}

The ``observed'' TOD streams used in this study were generated by
computer simulations that used the Level S software (Reinecke et
al.~\cite{Rei05}). The correspondence between the sample sequence of
the TOD and locations on the sky is determined by the scanning
strategy. The {\sc Planck} satellite will be placed in an orbit
around the 2$^{\rm nd}$ Lagrangian point (L2) of the Earth-Sun system
(Dupac \& Tauber~\cite{Dup05}). A satellite placed around L2 will
stay near the ecliptic plane and will follow the Earth when it is
orbiting the Sun.

The {\sc Planck} satellite rotates around its spin axis and the
angle between the spin axis and the optical axis of the telescope is
$85\degr$. While the satellite is spinning (at nominal rate 1 rpm)
the beam draws nearly great circles in the sky.  The satellite spin
axis is repointed at one-hour intervals.  The different scanning
strategies considered for {\sc Planck} (Dupac \&
Tauber~\cite{Dup05}) differ in what path these repointings follow on
the sky.  Between the repointings the spin axis remains fixed. We
applied a ``cycloidal'' scanning strategy where the spin axis
followed a circular path around the anti-solar direction.  The angle
between the spin axis and the Sun--Earth axis was $10\degr$, and the
spin axis completed a full circle around the Sun--Earth axis every 6
months (while the Sun--Earth axis itself of course made a full
circle along the ecliptic in one year).

The simulated TODs were generated for the originally planned 100 GHz
LFI detector number 9. The length of the TODs was 12 months. They
consisted of 525 960 scanning circles with 6498 samples on each
circle, corresponding to a sampling frequency of
$f_\mathrm{s}=108.3$~Hz. Since we assumed idealized satellite motion,
where the 60 scan circles between repointings fell exactly on each
other, sample by sample, these circles could easily be averaged into
a single ring.  This coadding of the TOD was performed before
destriping was applied, but it was not done for the ML codes.

In this study we utilized the
HEALPix\footnote{http://healpix.jpl.nasa.gov} pixelisation scheme.
Its pixel dimension is set by the $N_{\rm side}$ resolution
parameter. A map of the full sky contains $12N_{\rm side}^2$ pixels.

\begin{figure}
\begin{center}
\includegraphics[width=5.3cm,height=8.8cm,angle=90]{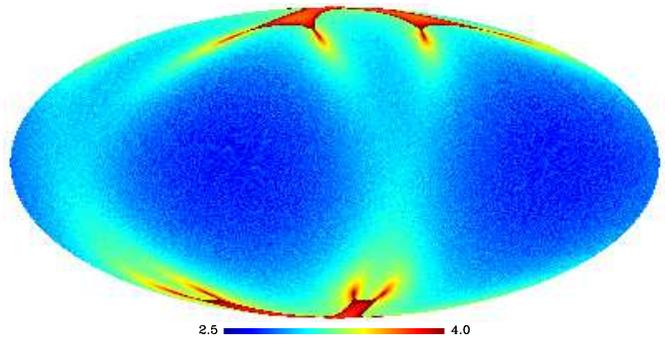}
\end{center}
\caption{Number of hits per pixel for the scanning strategy applied
in this study. The map is in the ecliptic coordinate system. The
scale is $\log_{10}(\rm{n_{\rm{hit}}})$, where $\rm{n_{\rm{hit}}}$
is the number of hits in a pixel. (A version of the paper with a
better-quality figure can be found at
http://www.physics.helsinki.fi/$\sim$tfo$\_$cosm/tfo$\_$planck.html.)}
\label{hits}
\end{figure}

The number of hits per pixel ($N_{\rm side} = 512$) of the applied
scanning strategy is shown in Fig.~\ref{hits}. At this resolution the
sky coverage was 100 \% (every pixel was hit).

The instrument noise was a sum of white and $1/f$ noise. The PSD of
the noise was \beq
   P(f) = \left(1+\frac{\fk}{f}\right)\frac{\sigma^2}{f_\mathrm{s}},
          \quad(f>f_\mathrm{min}). \label{psd}
\eeq Here $\fk$ is the knee frequency where the spectral powers of
the $1/f$ and white noise are equal. The nominal white noise
standard deviation (std) per integration time ($t =
1/f_{\mathrm{s}}$) is $\sigma$ and $f_{\rm min}$ is the minimum
frequency below which the noise spectrum becomes flat. The values of
the noise parameters are given in Table~\ref{parameters}. They
represented a realistic expected noise performance of the instrument
hardware. We used the stochastic differential equation (SDE)
algorithm (from Level S) to generate the TODs of the instrumental
noise. Perfect knowledge of the noise parameter values was assumed
both in the ML map-making and in the power spectrum estimation.

\begin{table}[ht!]
\caption[a]{\protect\small Simulation parameters used in the TOD
generation. Two TODs (signal-only and signal+noise) were generated
for all 4 simulation cases leading to 8 TODs in total. The TODs were
generated using Level S software (Reinecke et al.~\cite{Rei05}). The
TODs are sums of CMB (C), foreground (F), and/or noise (N) as
indicated in the table. The CMB and foreground TODs were made using
totalconvolver and interpolation algorithms (cases 1, 2 and 4) or
they were scanned from a high resolution map (case 3). This is also
indicated in the table.}
\begin{flushleft}
\begin{tabular}{lll}
\hline \hline Parameters common to all cases & &\\ \hline
Detector & & LFI 100~GHz \\
Number of detectors & & 1 \\
Mission time  & & 12 months  \\
Scanning (a) & & Cycloidal\\
Noise & &\\
  & $\sigma$ (b) & 3957.26~$\mu$K \\
  & $f_{\rm min}$ & $10^{-4}$~Hz \\
  & $f_{\rm s}$ & 108.3~Hz \\
\end{tabular}
\end{flushleft}
\begin{center}
\begin{tabular}{lllll}
Parameters for   & Case 1  & Case 2   & Case 3  & Case 4 \\
\hline
TOD (S only) & C+F (c) & C+F (c) & C (d) & C (c)\\
TOD (S+N) & C+F+N & C+F+N & C+N & C+N\\
Beam (e,f) & Symmetric & Elliptic & Symmetric & Elliptic \\
Noise - $\fk$ & 0.03~Hz & 0.03~Hz & 0.1~Hz & 0.1~Hz\\
\hline
\end{tabular}
\end{center}
\begin{tabular}{l}
(a) $10\degr$ amplitude and 6 months period.\\
(b) White noise std of the detector TOD (in antenna temp. units).\\
(c) Signal made using totalconvolver and interpolation.\\
(d) Signal scanned from a high resolution map.\\
(e) Symmetric Gaussian beam:\\Full width half maximum (FWHM) = 10.6551~arcmin.\\
(f) Elliptic Gaussian beam:\\FWHM$_{\rm major}$ = 11.8652~arcmin, FWHM$_{\rm minor}$ = 9.5684~arcmin.\\
\end{tabular}
\label{parameters}
\end{table}

We used 8 distinct TODs for this study. They were split in four
cases (2 TODs in a case). The simulation parameters for the TODs are
shown in Table~\ref{parameters}. The differences between the cases
were the knee frequencies of the instrument noise, the telescope
beams, whether the foreground\footnote{{\sc Planck} foreground
template maps were applied in the TOD generation. The templates are
available for the {\sc Planck} collaboration at
http://planck.mpa-garching.mpg.de} was included, and the ways how
the TODs were generated. The sky contained CMB and foreground
emissions in cases 1 and 2 but only CMB in cases 3 and 4. Four TODs
contained signal+noise and four TODs had signal only. The TODs were
convolved with either a symmetric or an elliptic instrument beam
(see Table~\ref{parameters}).

The CMB signal was derived from a set of $a_{\ell m}$ expansion
coefficients that was a realisation from a theoretical CMB angular
power spectrum $C_\ell^{\rm{th}}$. This $C_\ell^{\rm{th}}$ was
computed using the CMBFAST code\footnote{http://www.cmbfast.org}
(Seljak \& Zaldarriaga~\cite{Sel96}), and it corresponds to a
$\Lambda$CDM (cosmological constant + Cold Dark Matter) model.

For the cases 1, 2 and 4, the expansion coefficients of the sky were
convolved with the beam using the total convolution technique
(Wandelt \& G\'orski~\cite{Wan01}). The totalconvolver algorithm
(part of Level S) outputs a discrete temperature field which is
tabulated in an equally spaced three dimensional grid (two
dimensions for the pointing of the beam centre and one dimension for
the beam orientation). The TOD samples were interpolated from the
tabulated temperature grid. For the case 3, the signal TOD was
scanned from a high resolution map generated from the $a_{\ell m}$
that had been convolved with the beam. The map had $N_{\rm side} =
1024$ and it was generated with the SYNFAST code of the HEALPix
package.

\section{Output maps}
\label{sec:maps}

We had 8 TODs available for map-making comprising 4 cases (a
signal+noise TOD and a signal-only TOD for each case). Output maps
were made from these TODs using three map-making codes, leading to
24 output maps in total.

The 4 cases differed in a number of ways: in whether foreground was
included, in noise, in beam shape, and in how TODs were generated.
These differences represent both real effects and imperfections
(e.g., interpolation errors) in TOD generation.  They result in
variations of map-making performance from case to case. However,
these variations are not the object of this paper.  (The TODs
differed in too many ways for a study of the effect of these
differences to be conducted from just 4 cases).  Instead, the object
is to compare map-making methods. We have included all 4 cases
available to us, mainly to see whether the results of comparison
between methods stay consistent from case to case.  They do.  Some
conclusions regarding the effect of foregrounds on the different
map-making methods can also be drawn.

All output maps were made with pixel resolution $N_{\rm side} = 512$.
The noise in the signal+noise TODs resembled the noise of a single
LFI 100~GHz detector. Thus the maps are much noisier than they would
be if they were made from a full set of 24 detectors, by a factor of
about $\sqrt{24}$.

Before the ROMA and MapCUMBA output maps could be made the noise
filters had to be produced (see Sect.~\ref{subsec:igls}). They were
determined from the analytical model of the noise PSD (see Eq.
(\ref{psd})) and its known parameter values. The noise filters were
symmetric and had $N_{\xi} = 65537$ elements at non-negative lags
(lag $\ge 0$). The applied noise filters were identical in ROMA and
in MapCUMBA. The conjugate gradient iterations were continued until
the fractional difference had decreased to $< 10^{-6}$.

For destriping the amount of TOD was reduced by averaging 60 scan
circles between the repointings. The baseline amplitudes were solved
exactly (no iterations) from Eq. (\ref{destr2}) using as an
additional condition that their sum is zero.

\begin{figure}
\begin{center}
\includegraphics[width=5.3cm,height=8.8cm,angle=90]{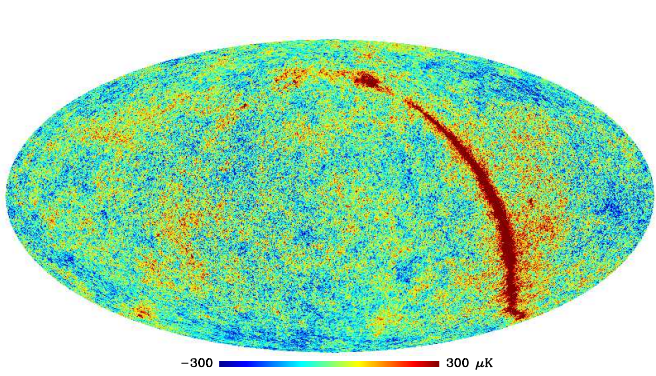}
\end{center}
\caption{The ROMA output map for case 1. The units are antenna
Kelvins at 100~GHz. The maps for MapCUMBA and destriping look
similar. This map is an output from one detector. For this plot the
map resolution was degraded to $N_{\rm side} = 256$. The monopole was
removed from the map.} \label{map_igls1}
\end{figure}

\begin{figure}
\begin{center}
\includegraphics[width=5.3cm,height=8.8cm,angle=90]{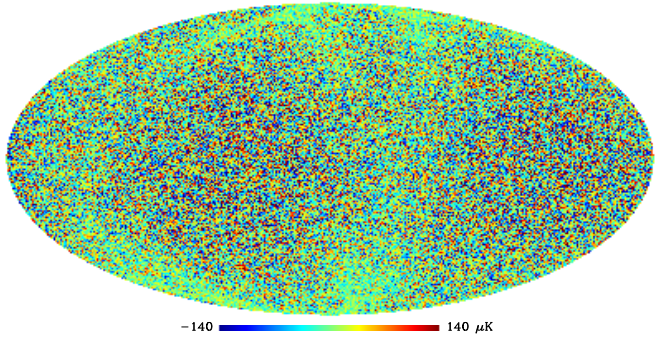}
\end{center}
\caption{The ROMA reconstruction error map for case 1. The units are
antenna Kelvins. The maps for MapCUMBA and destriping look similar.
The map is an output from one detector. For this plot the map
resolution was degraded to $N_{\rm side} = 256$. (A version of the
paper with a better-quality figure can be found at
http://www.physics.helsinki.fi/$\sim$tfo$\_$cosm/tfo$\_$planck.html.)}
\label{map_igls2}
\end{figure}

\begin{figure}
\begin{center}
\includegraphics[width=5.3cm,height=8.8cm,angle=90]{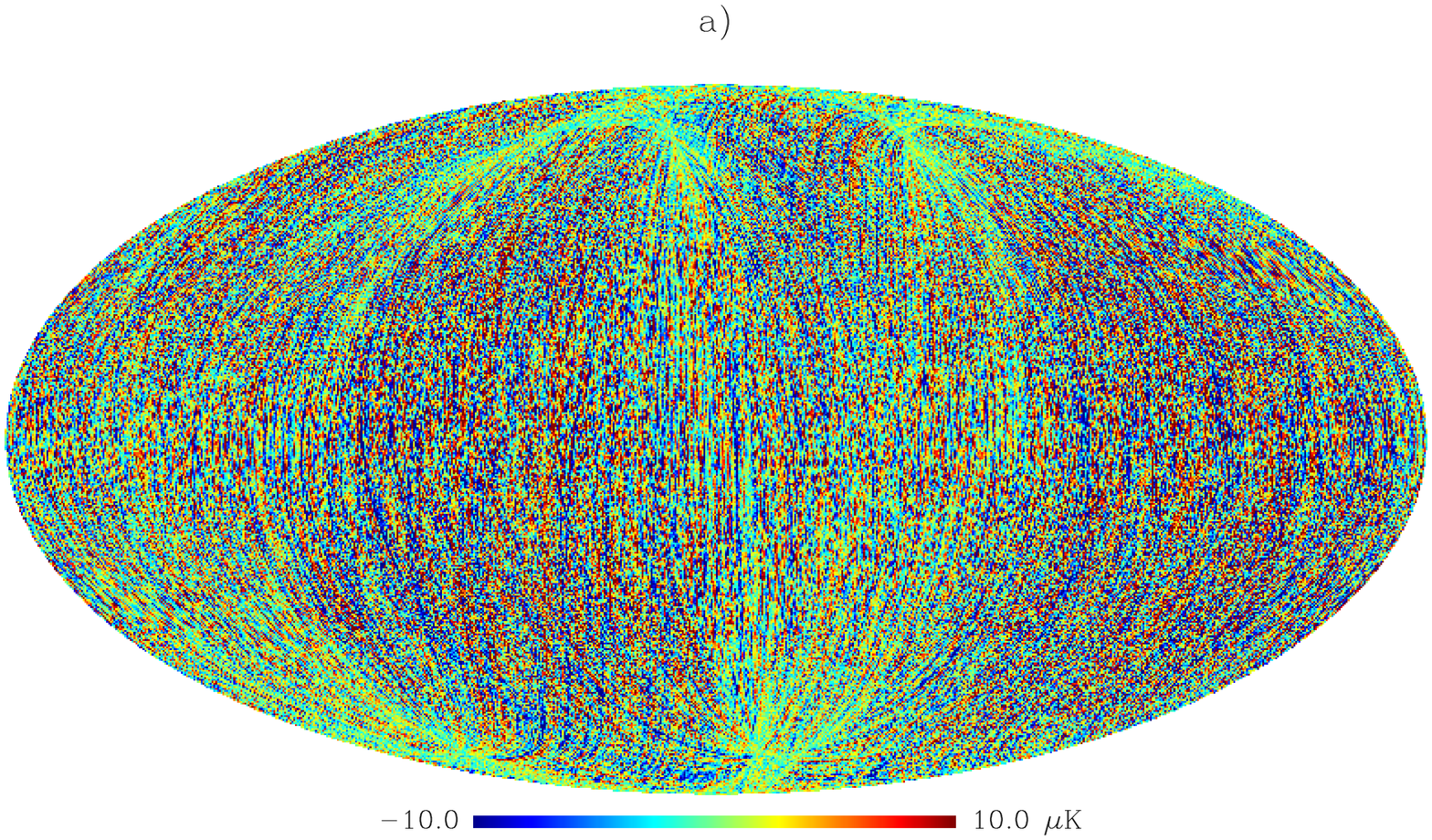}
\includegraphics[width=5.3cm,height=8.8cm,angle=90]{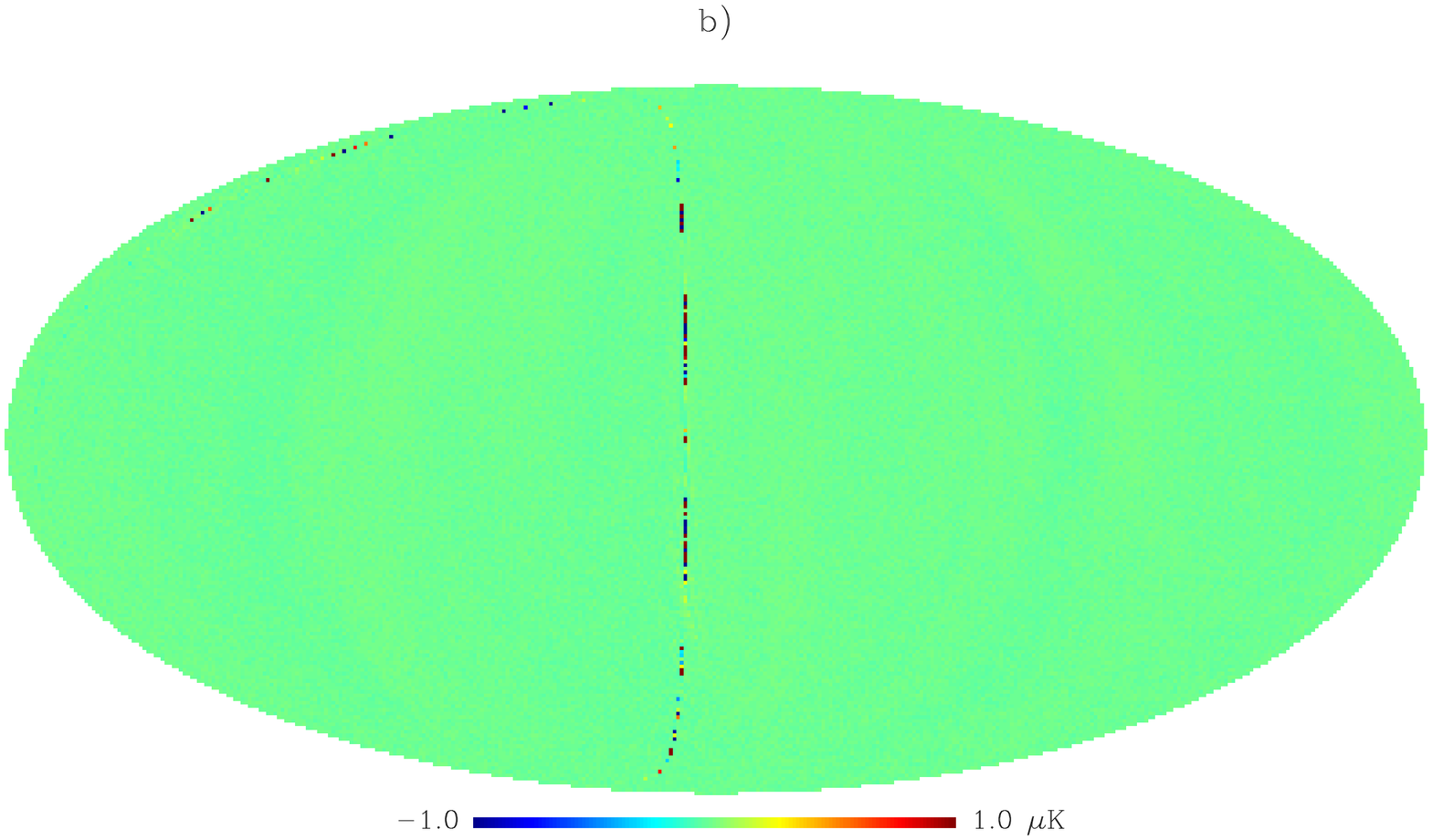}
\end{center}
\caption{{\bf a)} Difference map between the output maps of ROMA and
destriping. {\bf b)} Difference map between the output maps of ROMA
and MapCUMBA. The pixel minimum, maximum and std values (from the
$N_{\rm side} = 512$ maps) for map a) are (-109.7, 61.1, 10.0914)
$\mu$K and for map b) (-27.4, 38.7, 0.1531) $\mu$K. The units are
antenna Kelvins. All maps are for case 1. The corresponding
difference map between MapCUMBA and destriping looks similar to map
a). For the plots the map resolution was degraded to $N_{\rm side} =
256$. (A version of the paper with a better-quality figure can be
found at
http://www.physics.helsinki.fi/$\sim$tfo$\_$cosm/tfo$\_$planck.html.)}
\label{diff_igls_destr_mapcumba}
\end{figure}

The output maps of ML map-making and destriping can be divided into
a {\it binned noiseless map} and a {\it reconstruction error map}
(Tegmark~\cite{Teg97a}). The binned noiseless map is the signal-only
TOD binned to map pixels. The reconstruction error map ($\beps$) is
the (unwanted) deviation of the output map from the binned noiseless
map. A goal of map-making is to minimize this error. The
reconstruction error map can be further split into the signal
($\beps_p$) and noise ($\beps_n$) components: $\beps = \beps_p +
\beps_n$. These are discussed in detail in
Appendix~\ref{sec:recerror}. As shown there, the signal component
arises from pixelisation noise (D\'ore et al.~\cite{Dor01}). The
noise component ($\beps_n$) is the output map from the noise part of
the TOD.

Note that the map-making methods are linear (up to numerical
accuracy effects), so that these components can be obtained
separately and studied independently, by applying the codes to the
noise and signal parts of the simulated TODs (see e.g. de Gasperis
et al.~\cite{deG05}).

Our maps express the temperature fluctuations in the antenna
temperature. The ratio of the thermodynamic temperature fluctuation
to the antenna temperature fluctuation is $(e^x-1)^2/x^2e^x$, where
$x = h\nu/kT_{\rm{0}}$, $h$ is the Planck constant, $\nu$ is the
frequency, $k$ is the Boltzmann constant and $T_{\rm{0}}$ = 2.725 K
is the CMB temperature. For this study the ratio is 1.287 ($\nu$ =
100~GHz).

The output maps and the reconstruction error maps from all three
codes look similar.  Figs.~\ref{map_igls1} and~\ref{map_igls2} show
them for case 1. To see the difference, one needs to calculate the
difference map between the output maps of the different codes
(Fig.~\ref{diff_igls_destr_mapcumba}). The prominent feature in the
ROMA - MapCUMBA difference map (see
Fig.~\ref{diff_igls_destr_mapcumba}b) falls on top of the last
repointing period of the scan. In MapCUMBA the TOD is extended by
copying samples from its beginning to the end and vice versa. In
ROMA the TOD is extended by padding zeros to the end. The TOD
extension is required to make its length suitable for convolution
with the noise filter (number of TOD samples to be an integer
multiple of $2(N_\xi - 1)$), but the added samples are not projected
on the output map. The feature appearing in the ROMA - MapCUMBA
difference map reflects the different treatments of the end of the
TOD.

\begin{table}
\caption[a]{\protect\small The std, minimum and maximum values (all
in $\mu$K at 100~GHz antenna scale) of the pixel temperatures of the
reconstruction error maps ($\beps = \beps_p + \beps_n$). The numbers
were calculated from $N_{\rm side} = 512$ maps. They can be compared
to the std of a white noise map, calculated as $\sqrt{\langle 1/n
\rangle}\sigma$, where $\langle 1/n \rangle$ is the mean (taken over
the hit pixels) of the inverse of the number of hits in a pixel. The
values are 137.219 $\mu$K for cases 1 and 2 and 137.400 $\mu$K for
cases 3 and 4. The values are different between the cases due to
small differences in the cycloidal scannings.This white noise std
represents the level below which one cannot get. The difference
between ROMA and MapCUMBA is negligible. We give the numbers in the
table with many digits just to show at what level this difference
is. The numbers in this table and the next one should be compared in
the horizontal direction, to see the difference between the codes.
The differences in the vertical direction reflect the effect of
several differences in how the TODs were generated, but these
effects are not the object of this paper. }
\begin{center}
\begin{tabular}{llll}
\hline \hline
 std for $\beps$ \\(min, max)   &  ROMA     &  MapCUMBA   &   Destriping   \\
\hline
Case 1        & 138.2497 & 138.2496 & 138.455 \\
                 & (-825.3, 876.0) & (-825.3, 876.0) & (-822.5, 891.1) \\
Case 2         & 138.2475 & 138.2474 & 138.454 \\
                 & (-825.0, 876.7) & (-825.0, 876.7) & (-822.5, 891.2) \\
                 \hline
Case 3        & 138.9109 & 138.9110 & 139.397 \\
                 & (-838.4, 938.2) & (-838.1, 938.1) & (-851.2, 952.3) \\
Case 4         & 138.9114 & 138.9114 & 139.398 \\
                 & (-838.1, 937.9) & (-838.1, 937.9) & (-851.2, 952.2) \\
\hline
\end{tabular}
\end{center} \label{stds_sn}
\end{table}

\begin{table}
\caption[a]{\protect\small Same as Table~\ref{stds_sn} but now the
values are for the signal component of the reconstruction error map
($\beps_p$). The units are antenna $\mu$K (at 100~GHz).}
\begin{center}
\begin{tabular}{llll}
\hline \hline
std for $\beps_p$ \\(min, max)    &  ROMA   & MapCUMBA  &  Destriping  \\
\hline
Case 1        & 0.8794 & 0.8796 & 0.281 \\
                 & (-100.7, 53.8) & (-100.7, 53.8) &(-3.0, 2.1) \\
Case 2         & 0.6140 & 0.6142 & 0.252 \\
                 & (-62.2, 37.8) & (-62.2, 37.8) & (-2.3, 1.8) \\
\hline
Case 3        & 0.3130 & 0.3130 & 0.120 \\
                 & (-2.0, 1.9) &(-2.0, 1.9) & (-0.6, 0.7) \\
Case 4         & 0.4064 & 0.4064 & 0.169 \\
                 & (-2.8, 3.1) & (-2.8, 3.1) & (-0.9, 0.8) \\
\hline
\end{tabular}
\end{center} \label{stds_s}
\end{table}

Reconstruction error maps ($\beps$) were made by subtracting the
binned noiseless map from the signal+noise output maps. Minimum,
maximum and std values of the pixel temperatures of the
reconstruction error maps are given in Table~\ref{stds_sn}. In terms
of the map variance ML map-making is slightly better than
destriping. Table~\ref{stds_sn} also shows that the ML std is higher
than the std of the white noise indicating that excess noise remains
in the map. The cases 3 and 4 have higher map std's than the cases 1
and 2. This is mainly caused by the higher knee frequency of the
instrument noise in the cases 3 and 4. The map variances of ROMA and
MapCUMBA are practically equal showing that their performances are
similar. The results shown in Table~\ref{stds_sn} represent the
performance of a single LFI detector. For frequency maps made from
the observations of multiple LFI detectors the noise levels would be
correspondingly lower. For comparison it can be noted that the noise
level of a one-year W-band (94~GHz) frequency map ($N_{\rm side} =
512$) of the {\sc WMAP}\footnote{http://map.gsfc.nasa.gov}
experiment (from a total of eight W-band detectors making up 4
differencing assemblies) is slightly higher ($\sim$142 $\mu$K) than
the noise levels of Table~\ref{stds_sn}, which represents just a
single LFI detector.

\begin{figure}
\begin{center}
\includegraphics[width=5.3cm,height=8.8cm,angle=90]{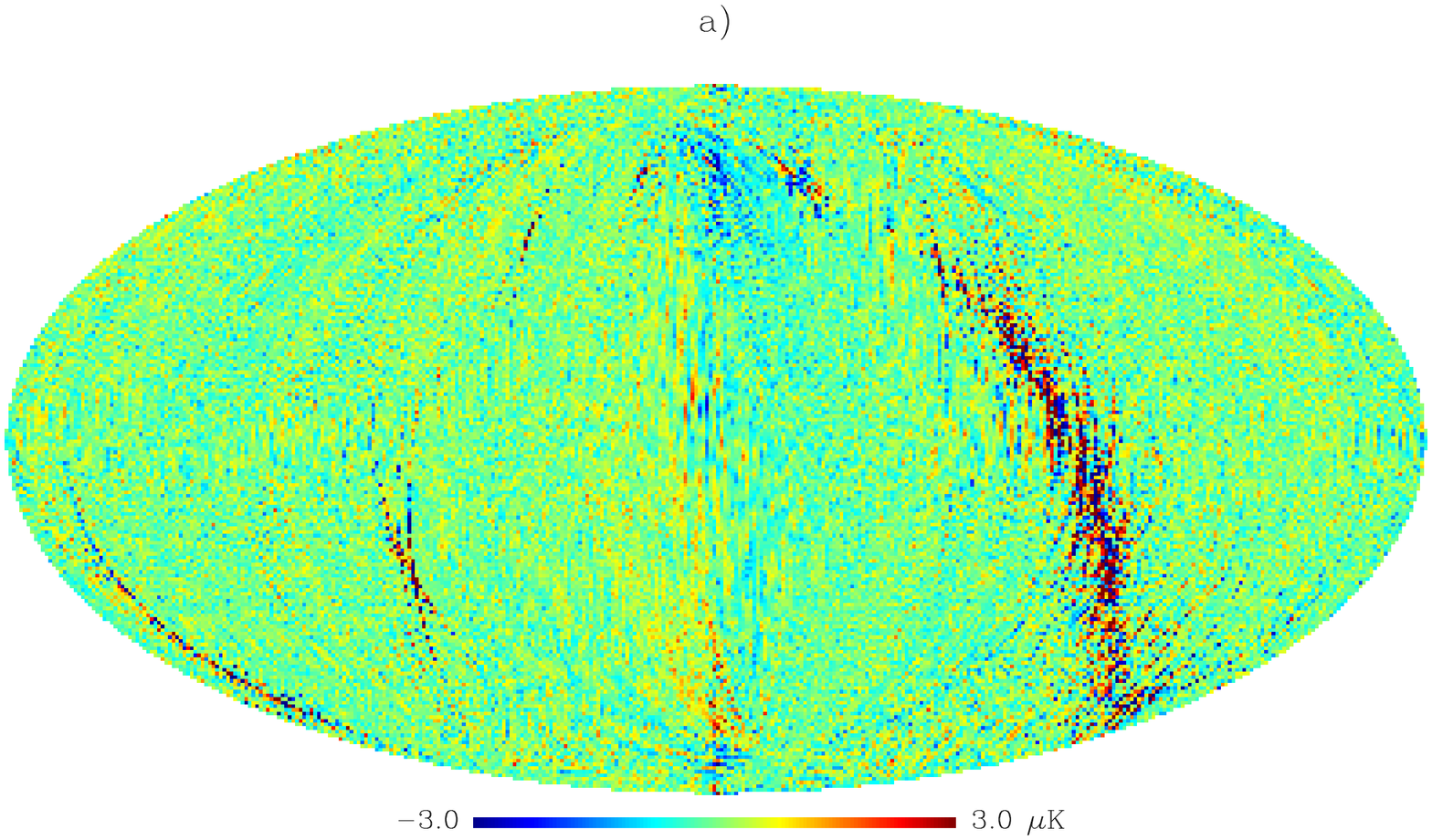}
\includegraphics[width=5.3cm,height=8.8cm,angle=90]{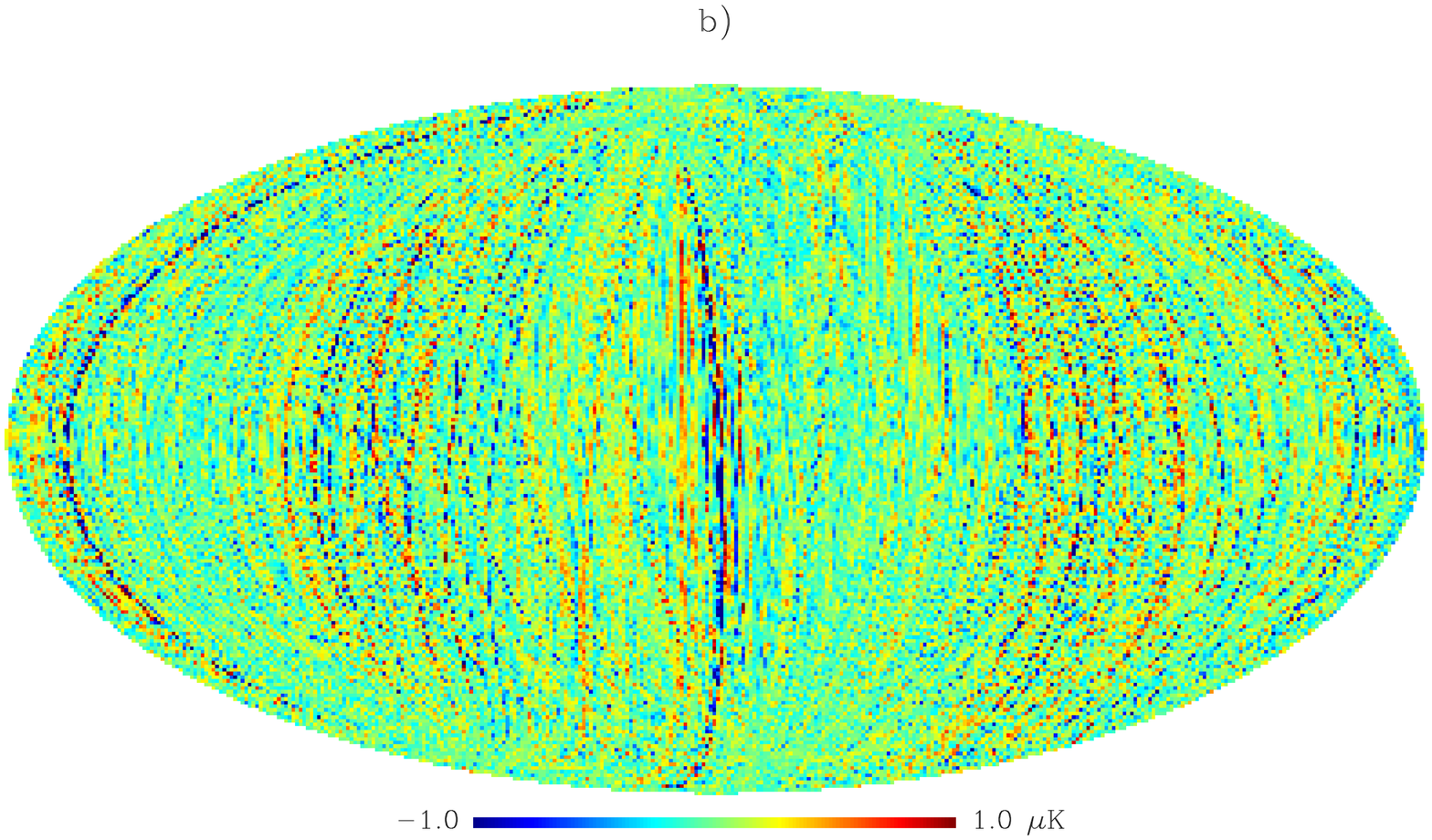}
\end{center}
\caption{The signal components of the reconstruction error maps for
ROMA ({\bf a}) and for destriping ({\bf b}). These are difference
maps between the signal-only output maps and the binned noiseless
maps. The output maps contained CMB and foreground. The units are
antenna Kelvins. All maps are for case 1. The corresponding map for
MapCUMBA looks similar to map a). The map resolution of the plots is
$N_{\rm side} = 512$. Note the different colour scales of the maps.
(A version of the paper with a better-quality figure can be found at
http://www.physics.helsinki.fi/$\sim$tfo$\_$cosm/tfo$\_$planck.html.)}
\label{residual_pixnoise}
\end{figure}

Pixel statistics for the signal components of the reconstruction
error maps ($\beps_p$) are given in Table~\ref{stds_s}. These maps
were produced by subtracting the binned noiseless map from the
noiseless signal-only output maps. In ML map-making the TOD was
convolved with the noise filter. As an example, the signal
components of the reconstruction error maps for case 1 are shown in
Fig.~\ref{residual_pixnoise}. The maps contained CMB and foreground
signals. The error magnitudes are larger for the ML map-making than
for the destriping. Additionally, the largest errors in the ML map
tend to locate where the foreground signal (actually, its gradient)
is strongest. For destriping this error appears as (erroneous)
baselines corresponding to offsetting an entire ring.  Therefore a
similar correlation between the errors and the foreground signal is
not visible in the destriped maps, although we expect that the
largest errors still originate from where the foreground signal is
strong.

The methods that were used to generate the signal parts of the TODs
were not uniform (total convolution vs. scanning of a high resolution
map, see Sect.~\ref{sec:tod}). That complicates the comparison of the
results of Table~\ref{stds_s} between the beams. This comparison was
not attempted in this study. However, the results can well be
compared (at a given beam) between different map-making algorithms
which was the main purpose of this study.

Table~\ref{stds_s} shows that the signal-only ML maps deviate more
from the binned noiseless map than what happens in destriping.
Foreground increases this error and large errors may occur in some
pixels of the ML maps. The std of $\beps_p$ is small compared to the
std of the overall reconstruction error map (see Table~\ref{stds_sn})
which is an indication that the total error is dominated by the
instrument noise.

The source of $ \beps_p$ is the pixelisation noise (as shown in
Appendix~\ref{sec:recerror}). In ML map-making the pixelisation
noise spectrum up to the knee frequency of the instrument noise
contributes to $ \beps_p$. In destriping, however, only a lower
frequency part contributes leading to a smaller $ \beps_p$. This is
addressed in more detail in Appendix~\ref{sec:recerror}. The
galactic foreground signal has a stronger spatial variation than
CMB. This results in higher pixelisation noise, which explains the
higher $ \beps_p$ magnitudes in the cases 1 and 2 than in the cases
3 and 4 (see Table~\ref{stds_s}).

\begin{figure}[tbh]
\center{\resizebox{\hsize}{!}{\includegraphics[width=4cm,height=3cm]{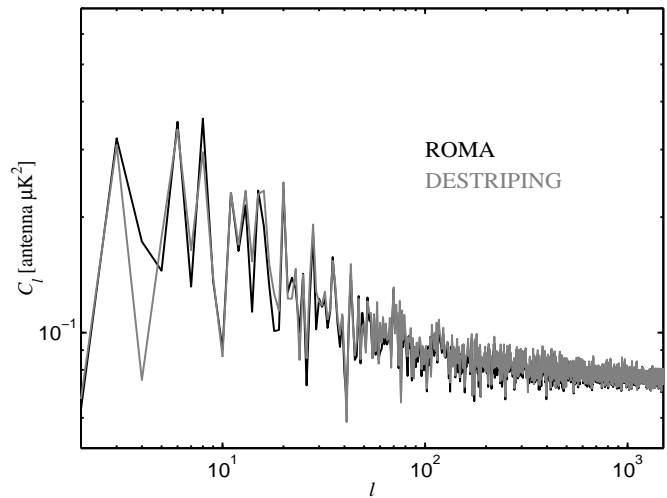}}}
\caption{The angular power spectra of the reconstruction error maps
($ \beps$) for ROMA and destriping. The angular power spectra for
ROMA and MapCUMBA were very similar and would be on top of each
other in this plot. The curves are for case 4.}
\label{recerror_spec1}
\end{figure}

\begin{figure}[tbh]
\center{\resizebox{\hsize}{!}{\includegraphics{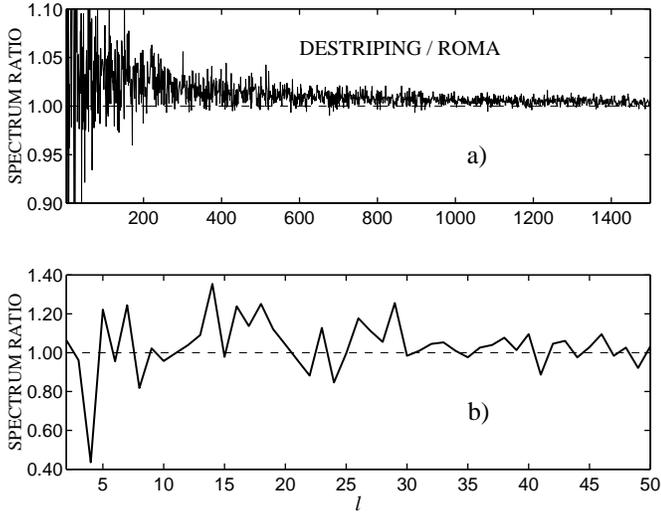}}}
\caption{The ratio of the angular power spectra of
Fig.~\ref{recerror_spec1}. {\bf a)} Full curve. {\bf b)} Zoom to low
$\ell$.} \label{recerror_spec2}
\end{figure}

We calculated the angular power spectra of the reconstruction error
maps ($ \beps$) (Fig.~\ref{recerror_spec1}). The spectra of ROMA and
MapCUMBA were very similar and could not be distinguished in the
plot. The ratio of the power spectra between destriping and ROMA is
shown in Fig.~\ref{recerror_spec2}. It seems that in destriping $
\beps$ has higher power in most multipoles.

To examine this further, 100 MC noise-only TODs were produced from
the known PSD of the instrument noise (see Eq. (\ref{psd})). The MC
noise TODs were generated by the SDE method using a different seed
value for every realisation. To have a reasonable calculation time
for this MC study, multiple 35~hour chunks of noise were generated
simultaneously in parallel processing and the chunks were glued one
after another at the end. This leads to an MC noise TOD with no
correlation between the chunks. Because the correlation of the noise
in the observed TODs is weak at $>$ 35~hour
lags\footnote{$R(\tau)/R(0) \approx 4\times10^{-5}$, where $R(t)$ is
the autocorrelation of the instrument noise (an inverse Fourier
transform of the noise PSD Eq. (\ref{psd})) and $\tau$ = 35~hours.},
the zero correlation beyond 35~hours in the MC noise is expected to
cause an insignificant error in the noise bias estimates.

\begin{figure}[tbh]
\center{\resizebox{\hsize}{!}{\includegraphics[width=4cm,height=3cm]{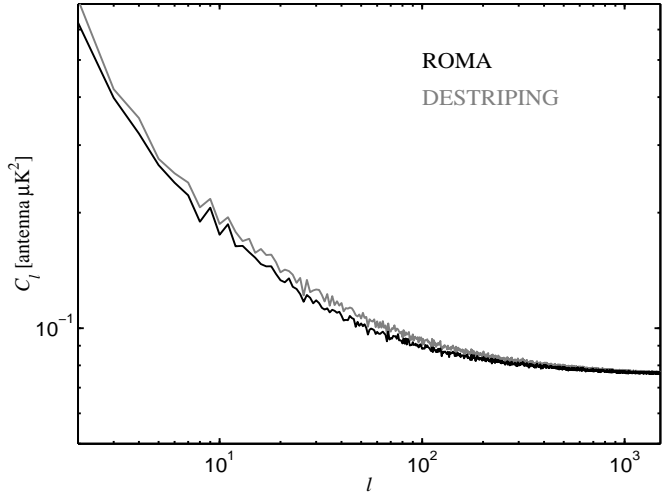}}}
\caption{Mean angular power spectra of output maps from 100 MC
realisations of noise TODs. The realisations were derived from the
known PSD of the instrument noise (see Eq. (\ref{psd})). Because the
MC TODs were noise-only, this plot is for cases 3 and 4.}
\label{noisebias}
\end{figure}

Output maps for ROMA and destriping were made from the MC noise TODs
and their angular spectra were derived. The mean spectra are shown
in Fig.~\ref{noisebias}. It shows that the mean angular power is
higher in destriping at all scales. The lower power at some $\ell$
(in Fig.~\ref{recerror_spec2}) seems to be just due to random
variation.

The map-making codes were run on an IBM SP RS/6000 computer with a
cluster of Power3 processors running at a clock speed of 375~MHz.
ROMA and MapCUMBA codes were run parallel in multiple processors
(number of processors was typically between 192 and 256) and it took
$\sim$10~min to produce an ML output map. In destriping an output
map was produced in $\sim$7~min in a single processor job. In MC
studies this time can be reduced to $\sim$4~min by inverting the
matrix (see Eq. (\ref{destr2})) once and using the inverse in the
subsequent runs.

Note that a part of this large difference in computation cost is due
to destriping being applied to a coadded TOD, which was a factor 60
shorter than a full TOD.  This coadding was possible without error,
because we assumed idealized pointing.  In reality, the pointings of
the different circles of the same ring do not fall exactly on top of
each other.  This means that either some additional error is
introduced by the coadding, or that destriping has to work with the
actual pointings of the full TOD.  The latter option increases the
computational cost of destriping, but it will still be significantly
less than for ML map-making.

\section{Power spectrum estimates}
\label{sec:estimates}

\begin{figure}[tbh]
\center{\resizebox{\hsize}{!}{\includegraphics[width=7.5cm,height=6cm]{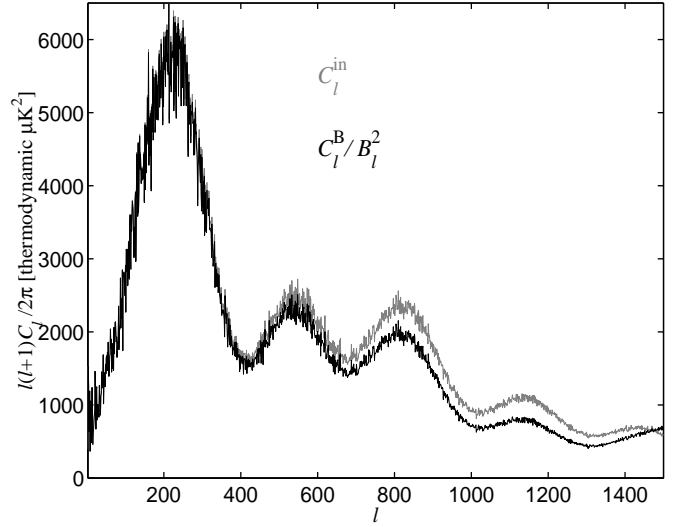}}}
\caption{The angular power spectrum ($C_{\ell}^{\rm B}$) of the
binned noiseless map (black curve). The map ($N_{\rm side} = 512$)
was binned from the simulated noiseless TOD containing CMB that was
smoothed with the symmetric beam (case 3, see
Table~\ref{parameters}). The $C_{\ell}^{\rm B}$ spectrum shown in
the plot is deconvolved with the same symmetric beam response. For
comparison, the input power spectrum $C_\ell^{\rm{in}}$ of the CMB
sky is shown as well (grey curve). The difference between
$C_{\ell}^{\rm B}$ and $C_\ell^{\rm{in}}$ is (in this case) mainly
due to pixel window smoothing and is discussed in Appendix
\ref{sec:inputmap}.} \label{clb}
\end{figure}

To see how the differences between destriping and ML map-making are
reflected in the angular power spectrum estimates, we derived
$C_\ell$ estimates from the output maps of case 3, where the TOD
contained CMB and noise. Since the output maps from ROMA and
MapCUMBA were practically identical, we let ROMA represent ML
map-making in this section. The CMB was convolved with the symmetric
beam (case 3, see Table~\ref{parameters}). As defined in
Sect.~\ref{subsec:powerspectrum}, our power spectrum estimates
$\widehat{C}_{\ell}^{\rm B}$ are estimates of the angular power
spectrum of the binned noiseless map. The estimates were compared to
the actual spectrum ($C_{\ell}^{\rm B}$) of the binned noiseless
map. That map was binned from the noiseless TOD containing CMB only.
The angular spectrum $C_{\ell}^{\rm B}$ is shown in Fig.~\ref{clb},
which also shows the input spectrum $C_\ell^{\rm in}$. The
difference between $C_{\ell}^{\rm B}$ and $C_\ell^{\rm in}$ is due
to a number of effects, which are discussed in
Appendix~\ref{sec:inputmap}; but they are not relevant for our
comparison between map-making methods, as they can only produce
estimates of $C_{\ell}^{\rm B}$.

\begin{figure}[tbh]
\center{\resizebox{\hsize}{!}{\includegraphics[width=7.5cm,height=6cm]{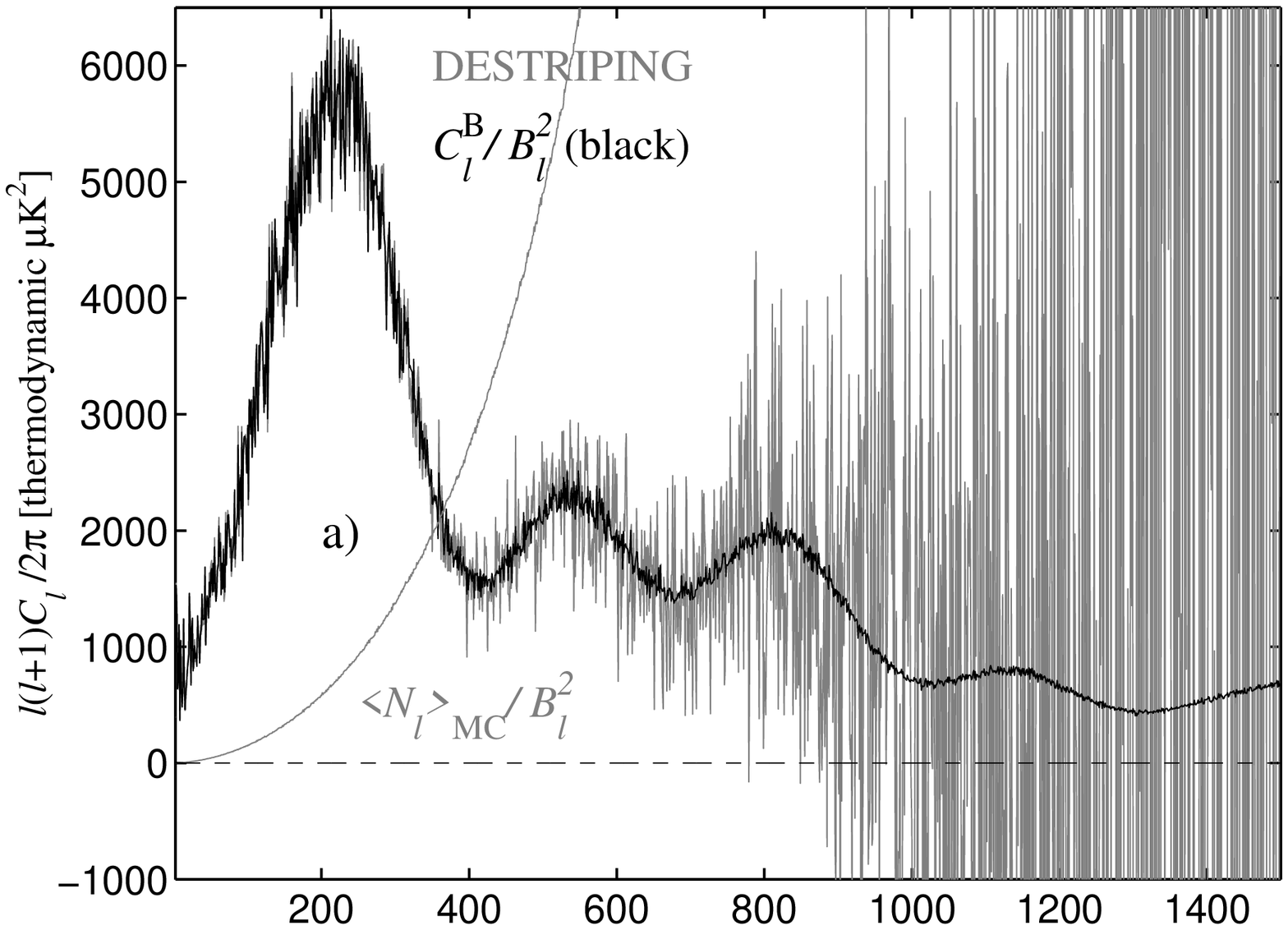}}}
\center{\resizebox{\hsize}{!}{\includegraphics[width=7.5cm,height=6cm]{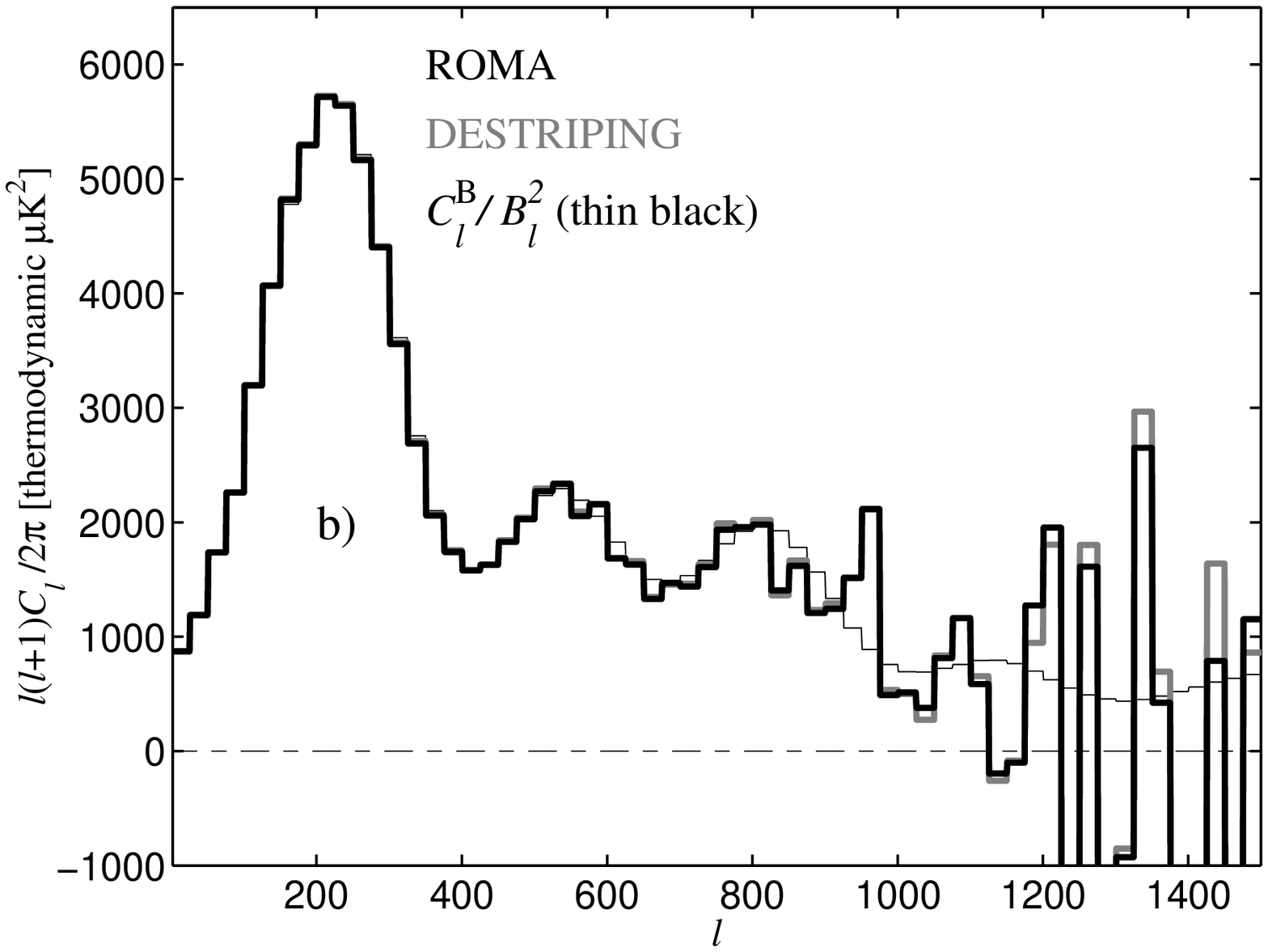}}}
\caption{{\bf a)} Angular power spectrum estimate
$\widehat{C}_{\ell}^{\rm B}$ for destriping (grey curve). The ROMA
estimate would fall nearly on top of the destriping estimate and
would not distinguish in this plot. The estimates were derived from
a single sky realisation (case 3). The output maps covered the full
sky and contained CMB and instrument noise.  Since just a single
detector was considered,  the noise becomes dominant already at
around $\ell \simeq 350$. $C_\ell^{\rm{B}}$ is the angular spectrum
of the binned noiseless map and $\langle N_\ell \rangle_{\rm{MC}}$
is the MC noise bias (for destriping). Note that they have been
deconvolved with a symmetric beam that was identical to the
instrument beam. The filter function $F_\ell$ was set to one for
both estimates. {\bf b)} Same as a) but the spectra have been $\ell$
binned to $\Delta\ell = 25$. The difference between the ROMA and
destriping estimates is visible at $\ell > 1000$.}
\label{spectrum_estimates}
\end{figure}

The relation between the pseudo spectrum $\widetilde{C}_\ell$
(angular power spectrum obtained from the output map) and the power
spectrum estimate ($\widehat{C}_{\ell}^{\rm B}$) was given in Eq.
(\ref{aps2}). The estimate is obtained by inverting the equation. The
estimate of the noise bias $\langle \widetilde{N}_\ell \rangle$ was
obtained from the MC simulations (see Fig.~\ref{noisebias}). The
value of $F_\ell$ was initially set to one. The obtained power
spectrum estimates, the spectrum of the binned noiseless map and the
MC noise bias (for destriping) are shown in
Fig.~\ref{spectrum_estimates}.

For the quality of an angular power spectrum estimate its bias and
covariance matrix are important figures of merit. For the covariance
matrix we restricted our study to its diagonal elements, which
represent the error bars of the estimate.

\begin{figure}[tbh]
\center{\resizebox{\hsize}{!}{\includegraphics{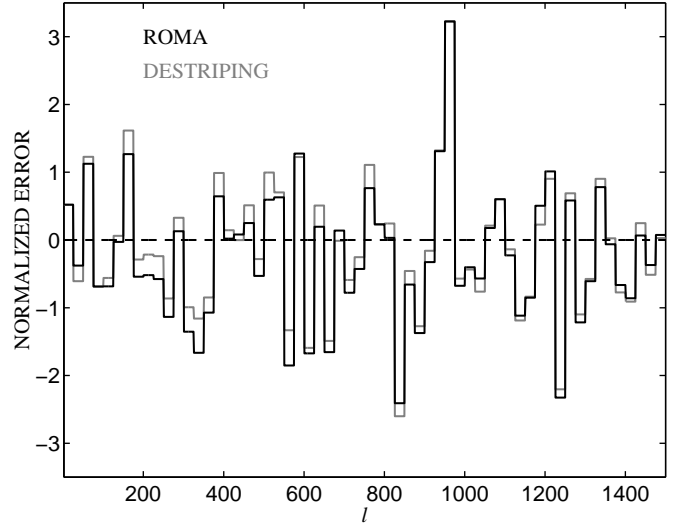}}}
\caption{The normalized power spectrum estimation errors for ROMA
and destriping. The normalized error was obtained by dividing the
binned estimation error (Eq. (\ref{delta_cb}), $\Delta\ell = 25$)
with the analytic approximation of its std (Eq. (\ref{stdcb})). The
std of destriping was used in all normalizations. The filter
function had value 1.0 for both curves. The bin-to-bin fluctuations
are mainly caused by the instrument noise.} \label{norm_error}
\end{figure}

\begin{figure}[tbh]
\center{\resizebox{\hsize}{!}{\includegraphics{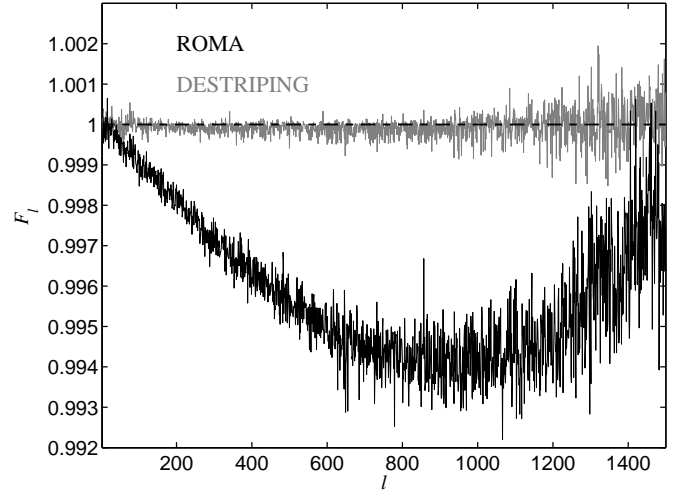}}}
\caption{Filter functions $F_\ell$ for ROMA and destriping. They
were obtained from Eq. (\ref{aps2}) (with $\langle
\widetilde{N}_\ell \rangle = 0$) by dividing the pseudo spectra of
the output maps of the noiseless CMB-only TOD with the spectrum of
the binned noiseless map. The filter function for MapCUMBA was
nearly identical to the filter function of ROMA and would not
distinguish from the ROMA filter function in this plot.}
\label{filter_function}
\end{figure}

\begin{figure}[tbh]
\center{\resizebox{\hsize}{!}{\includegraphics{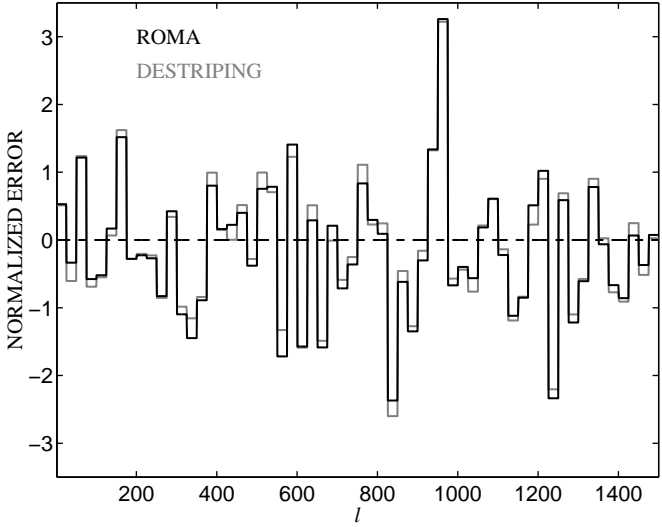}}}
\caption{Same as Fig.~\ref{norm_error} but now the power spectrum
estimates have been corrected with the estimates of the filter
function (from Fig.~\ref{filter_function}).}
\label{norm_error_filterfunction}
\end{figure}

\subsection{Bias}
\label{subsec:biaserror}

We defined the estimation error $\Delta \widehat{C}_\ell$ as the
difference between the power spectrum estimate and the spectrum of
the binned noiseless map: $\Delta \widehat{C}_\ell =
\widehat{C}_{\ell}^{\rm B} - C_\ell^{\rm{B}}$. The estimation error
was binned by averaging $\Delta \ell$ multipoles to a bin
 \beq
 \Delta \widehat{\mathcal{C}}_b = \frac{1}{\Delta \ell}
   \sum_{\ell \in b}{\frac{\ell(\ell + 1)\Delta \widehat{C}_\ell}{2\pi}}.
   \label{delta_cb}
 \eeq
We evaluated the binned errors $\Delta  \widehat{\mathcal{C}}_b$ for
ROMA and destriping from the spectra shown in
Fig.~\ref{spectrum_estimates}. To facilitate the comparison of the
binned estimation errors, we normalized them by dividing them with an
analytic approximation of their std, which was obtained from
 \beq
   \sigma_b = \left[\frac{1}{{\Delta \ell}^2}
   \sum_{\ell \in b}{\left(\frac{\ell(\ell +
   1)}{2\pi}\right)^2\sigma_\ell^2}\right]^{1/2},
 \label{stdcb}
 \eeq
where
 \beq \sigma_\ell =
 \sqrt{\frac{2}{(2\ell+1)f_{\rm{sky}}}\left(2C_\ell^{\rm{B}}\langle
 \widetilde{N}_\ell \rangle + \langle \widetilde{N}_\ell
 \rangle^2\right)}. \label{efs}
  \eeq
is the approximation for the std of the unbinned error $\Delta
\widehat{C}_\ell$ (Efstathiou~\cite{Efs05}). In this formula
$C_{\ell}^{\rm B}$ is a given signal (no cosmic variance). Sky
coverage fraction is $f_{\rm{sky}} = 1$ in our case.  For all
normalizations we used the $\langle \widetilde{N}_\ell
\rangle_{\rm{MC}}$ obtained for destriping in place of $\langle
\widetilde{N}_\ell \rangle$.

The normalized errors are shown in Fig.~\ref{norm_error}. Their std
(from one noise realisation to another) should be $\sim$1. If an
angular power spectrum estimate has a non-zero bias the mean of the
fluctuations of the normalized error will have a positive or
negative trend. In the case of zero bias the mean will be close to
zero. We can note that the normalized errors are different for ROMA
and destriping, the largest differences being at $\ell < 800$. The
differences are, however, smaller than the std of the errors.

We could expect that some bias could be introduced to our power
spectrum estimates because we neglected (by setting $F_\ell$ = 1 in
Eq. (\ref{aps2})) the error that the map-making causes to the CMB
signal. This is a reflection of the signal component $\beps_p$ of
the reconstruction error that we found in the map domain (see
Sect.~\ref{sec:maps}). To assess the level of the bias we estimated
$F_\ell$ for ROMA and for destriping. We carried out no MC
simulations to estimate them, but we determined them from Eq.
(\ref{aps2}) using the pseudo spectra $\widetilde{C}_\ell$ from the
output maps of the noiseless (CMB-only) TOD ($\langle
\widetilde{N}_\ell \rangle = 0$). Because the values of $F_\ell$ are
based on one CMB realisation only, these results should be taken as
indicative.  (In fact, we now fully correct for the effect of
$\beps_p$, since the filter function is derived from the same
realisation to which it is applied.  In reality, of course, the
signal-only TOD will not be available, and the filter function
should be evaluated as an expectation value. It will then remove
only the bias due to $\beps_p$.)

The obtained $F_\ell$ are shown in Fig.~\ref{filter_function}. For
destriping there is essentially no filter function ($F_\ell \sim 1$,
Poutanen et al.~\cite{Pou04}). For ROMA the deviation from 1 is
larger, showing its largest values at $\ell = 800 \ldots 1000$. If
not corrected, the map-making errors cause a bias in the ML spectrum
estimates whose maximum value in this case would be $\sim$0.6\% of
the magnitude of the CMB spectrum.

We corrected our angular power spectrum estimates with the obtained
$F_\ell$ and reproduced the normalized estimation errors. The result
is shown in Fig.~\ref{norm_error_filterfunction}. When comparing to
Fig.~\ref{norm_error} the improved match between the normalized
errors of ROMA and destriping can be noted (especially at $\ell <
600$). The remaining differences are mainly due to the differences
in the noise of the output maps of these two algorithms.

\subsection{Error bars}
\label{subsec:errorbars}

An error bar is defined here as the square root of the diagonal
element of the covariance matrix $\langle
\Delta\widehat{C}_{\ell}\Delta\widehat{C}_{\ell'}\rangle$ ($\pm
1\sigma$ error bar). The error bars can be derived either
analytically (Tegmark~\cite{Teg97b}; Efstathiou~\cite{Efs04}) or by
MC simulations (Hivon et al.~\cite{Hiv02}; Poutanen et
al.~\cite{Pou04}). In this study we did not do signal+noise MC
simulations to determine the error bars, but used instead an
approximation to compare ROMA and destriping. We used
$\sigma_{\ell}$ from Eq. (\ref{efs}) with a modification that takes
into account the different filter functions for different map-making
algorithms ($F_\ell$ assumed value 1.0 in Eq. (\ref{efs})) \beq
\sqrt{\langle(\Delta\widehat{C}_{\ell})^2\rangle} =
\sqrt{\frac{2}{(2\ell+1)}}\sqrt{\frac{2F_\ell C_\ell^{\rm{B}}\langle
\widetilde{N}_\ell \rangle_{\rm{MC}} + \langle \widetilde{N}_\ell
\rangle_{\rm{MC}}^2}{F_\ell^2}}. \label{efs_f} \eeq The spectrum
$C_{\ell}^{\rm B}$ represents here a given signal (no cosmic
variance).

\begin{figure}[tbh]
\center{\resizebox{\hsize}{!}{\includegraphics[width=4cm,height=3cm]{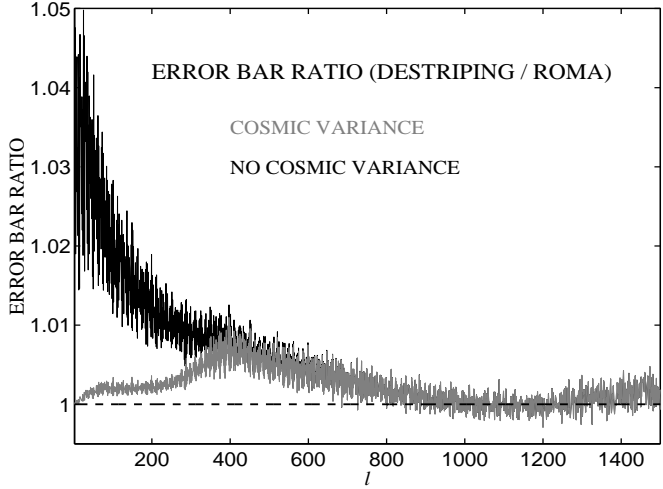}}}
\caption{An estimate for the ratio of the error bars between
destriping and ROMA. Black curve is for the estimation of a
particular CMB realisation of the sky (no cosmic variance, error
bars from Eq. (\ref{efs_f})) and the grey curve is for the
estimation of the underlying theoretical CMB spectrum (cosmic
variance included, error bars from Eq. (\ref{knox})).} \label{bars}
\end{figure}

Applying the spectrum $C_{\ell}^{\rm B}$ from Fig.~\ref{clb}, the
noise biases ($\langle \widetilde{N}_\ell \rangle_{\rm{MC}}$) from
Fig.~\ref{noisebias} and the filter functions from
Fig.~\ref{filter_function} the ratio of the error bars between
destriping and ROMA was evaluated. It is shown in Fig.~\ref{bars}
(black curve). The error bars at $\ell \lesssim 1000$ are larger for
destriping than for ROMA. The largest relative differences are
$\sim$5\%. The main cause of the larger error bars is the higher
level of noise in the output maps of destriping (see
Table~\ref{stds_sn}). At high-$\ell$ the larger map noise of the
destriping is partly compensated by its larger filter function (see
Fig.~\ref{filter_function}) leading to error bars that have nearly
the same magnitude as the ROMA error bars.

As a second case we assumed that we want to estimate the angular
spectrum $C_\ell^\mathrm{th}$ of the underlying theoretical CMB
(instead of its particular realisation as above). Cosmic variance
then increases the error bars (Scott et al.~\cite{Sco94}; Hobson \&
Magueijo~\cite{Hob96}):
 \beq
 \sqrt{\langle (\Delta
 \widehat{C}_\ell)^2 \rangle} = \sqrt{\frac{2}{(2\ell+1)}}\frac{F_\ell
 C_{\ell}^{\rm{B}} + \langle \widetilde{N}_\ell
 \rangle_{\rm{MC}}}{F_\ell}. \label{knox}
 \eeq
The ratio of these error bars is shown in Fig.~\ref{bars} as well
(grey curve). Because noise dominates the error bars at high
multipoles, these are similar to the error bars without cosmic
variance. Due to the dominance of cosmic variance at low multipoles,
the magnitudes of the error bars of the two methods approach each
other at low $\ell$.

\section{Conclusions}
\label{sec:conclu}

We have presented a comparison of the maps produced by three
different map-making codes and two map-making methods, destriping
and ML map-making. We also compared the angular power spectrum
estimates obtained from destriping and ML maps. The maps and power
spectra were derived from a set of one-year TOD streams that
resembled the observations expected from a single 100~GHz {\sc
Planck} LFI detector.

In terms of the map variance the two ML codes, ROMA and MapCUMBA,
produce nearly identical maps, with lower noise than destriping. This
lower noise is an advantage for them and it facilitates smaller error
bars for the power spectrum estimates. The difference is, however,
rather small.

ROMA and MapCUMBA require knowledge of the power spectrum of the
instrument noise, whereas destriping does not. In a real experiment
the noise spectrum (if required) needs to be estimated from the
observed data. Some estimation error can be expected which may
increase the noise in the ROMA and MapCUMBA maps. Thus differences
in the noise performance between ROMA/MapCUMBA and destriping may
become smaller in a real experiment. A perfectly known instrument
noise spectrum was assumed in this study.

The map-making methods caused errors (exhibited in the signal
component of the reconstruction error map) in the signal part of the
output maps. The origin of these errors is the sub-pixel structure of
the signal (pixelisation noise). It was shown that these errors were
smaller in destriping than in ROMA and MapCUMBA. It was further shown
that, if a proper correction is not applied, these errors may show up
as an extra bias in the power spectrum estimates. The extra bias
would be larger for ROMA and MapCUMBA than for destriping.

In terms of CPU resources destriping is less demanding. This is an
advantage in e.g. MC simulations.

\begin{acknowledgements}
The work reported in this paper was done by the CTP Working Group of
the {\sc Planck} Consortia. {\sc Planck} is a mission of the
European Space Agency. Some of this work was carried out in June
2002 and in January 2003 during two CTP meetings hosted by the
Institute of Astronomy (University of Cambridge). We thank their
hospitality during these meetings. This research used resources of
the National Energy Research Scientific Computing Center, which is
supported by the Office of Science of the U.S. Department of Energy
under Contract No. DE-AC03-76SF00098. This work has made use of the
{\sc Planck} satellite simulation package (Level S), which is
assembled by the Max Planck Institute for Astrophysics {\sc Planck}
Analysis Centre (MPAC). We acknowledge the use of the CMBFAST code
for the computation of the theoretical CMB angular power spectrum.
Some of the results in this paper have been derived using the
HEALPix package (G\'orski et al. \cite{Gor99},~\cite{Gor05a}). This
work was supported by the Academy of Finland grant no. 75065. TP
wishes to thank the V\"{a}is\"{a}l\"{a} Foundation for financial
support. The US {\sc Planck} Project is supported by the NASA
Science Mission Directorate.

\end{acknowledgements}


\appendix
\section{Reconstruction error map}
\label{sec:recerror}

In this Appendix we discuss in more detail the reconstruction error
map and how its signal and noise components arise.

The output map of the ML map-making can be solved from Eq.
(\ref{igls2}). That equation is reproduced here \beq
 \bP^T\binvN\bP\ve{m} = \bP^T\binvN\ve{y}. \label{app1}
\eeq

The noise covariance matrix $\bN$ can be freely normalized by a
constant without affecting the output map. Let us assume that each
element of $\bN$ has been divided by the variance ($\sigma^2$) of
the non-correlated noise component of the observed TOD (vector
$\ve{y}$). By replacing $\binvN$ with an identity $\bIt - (\bIt -
\binvN)$ and rearranging some of the terms one obtains for the
output map \beq
  \ve{m} = (\bP^T\bP)^{-1}\bB^{-1}\bP^T\binvN\ve{y}, \label{app2a}
\eeq where
\beq
  \bB = \bIm - \bP^T(\bIt-\binvN)\bP(\bP^T\bP)^{-1}. \label{app2b}
\eeq Here $\bIt$ and $\bIm$ are unit matrices with dimensions
equal to the number of samples $N_t$ of the TOD and the number of
pixels $N_{\rm{pix}}$ in the map, respectively. We can apply a
geometric series trick (Tegmark~\cite{Teg97a}) to prove the
following identity \bea
  [\bIm - \bP^T(\bIt-\binvN)\bP(\bP^T\bP)^{-1}]^{-1}\bP^T = \nonumber\\
          = \bP^T[\bIt - (\bIt-\binvN)\bP(\bP^T\bP)^{-1}\bP^T]^{-1}. \label{app3}
\eea Applying this in Eqs. (\ref{app2a}) and (\ref{app2b}) and
noting the definition of the matrix $\bZ$ (see
Sect.~\ref{subsec:destr}) one obtains for the output map \beq
  \ve{m} = (\bP^T\bP)^{-1}\bP^T[\bIt + (\bN - \bIt)\bZ]^{-1}\ve{y}. \label{app4}
\eeq By adding and subtracting $\ve{y}$ on the right hand side and
carrying out some arithmetic manipulations the output map can be
expressed in the following form \beq
  \ve{m} = (\bP^T\bP)^{-1}\bP^T[\ve{y} - \ve{\Delta}]. \label{app5}
\eeq
Vector $\ve{\Delta}$ is solved from the linear equation
\beq
  [\bZ + (\bN - \bIt)^{-1}]\ve{\Delta} = \bZ\ve{y}. \label{app6}
\eeq

Let the complex valued matrix $\bH$ ($[H]=(N_t,N_t)$) be an inverse
DFT (Discrete Fourier Transform) operator that converts frequency
domain vectors to time domain (to TOD domain): $\ve{y} =
\bH\widetilde{\ve{y}}$, where $\widetilde{\ve{y}}$ is the frequency
domain counterpart of $\ve{y}$. The inverse operator to $\bH$ is its
Hermitian conjugate $\bH^{\dag}$. We can assume that the matrix
$\bH$ is normalized to $\bH\bH^{\dag} = \bH^{\dag}\bH = \bIt$. A
square matrix $\bA$ ($[A]=(N_t,N_t)$) in the TOD domain can be
converted to a matrix $\widetilde{\bA}$ in the frequency domain:
$\widetilde{\bA} = \bH^{\dag}\bA\bH$.

After converting both sides of Eq. (\ref{app6}) into frequency
domain, the vector $\widetilde{\ve{\Delta}}$ (frequency domain
counterpart of $\ve{\Delta}$) can be solved from the equation \beq
  [\bH^{\dag}\bZ\bH + (\widetilde{\bN} - \bIt)^{-1}]\widetilde{\ve{\Delta}} = \bH^{\dag}\bZ\ve{y}. \label{app7}
\eeq The solution for the output map becomes then
 \beq
  \ve{m} = (\bP^T\bP)^{-1}\bP^T[\ve{y} - \bH\widetilde{\ve{\Delta}}]. \label{app8}
\eeq Comparing Eqs. (\ref{app7}) and (\ref{app8}) to the
corresponding Eqs. (\ref{destr2}) and (\ref{destr3}) of destriping
the similarity of the output map solutions between ML and destriping
can be clearly seen.

The observed TOD ($\ve{y}$) contains two components: signal $\ve{s}$
and instrument noise $\ve{n}$ (see Sect.~\ref{sec:tod}). The term
$\bZ\ve{y}$ (see Eq. ~(\ref{app7})) can now be split in two
components \beq
   \bZ\ve{y} = \bZ\ve{s} + \bZ\ve{n}. \label{app9}
\eeq
Writing out the first term on the right hand side we obtain
\beq
   \bZ\ve{s} = \ve{s} - \bP(\bP^T\bP)^{-1}\bP^T\ve{s}. \label{app10}
\eeq Apart from the sign, $\bZ\ve{s}$ is the pixelisation noise
introduced by Dor\'e et al. (\cite{Dor01}). Pixelisation noise
represents error that is caused by the discretization of the sky
into pixels. Following Dor\'e et al. (\cite{Dor01}) we define
pixelisation noise $\ve{p} = -\bZ\ve{s}$.

The split of $\bZ\ve{y}$ in two components means that
$\widetilde{\ve{\Delta}}$ is split in two components as well:
$\widetilde{\ve{\Delta}} =
\widetilde{\ve{\Delta}}_p+\widetilde{\ve{\Delta}}_n$. They can be
solved from \beq
  [\bH^{\dag}\bZ\bH + (\widetilde{\bN} - \bIt)^{-1}]\widetilde{\ve{\Delta}}_p = -\bH^{\dag}\ve{p} \label{app11}
\eeq
and
\beq
  [\bH^{\dag}\bZ\bH + (\widetilde{\bN} - \bIt)^{-1}]\widetilde{\ve{\Delta}}_n = \bH^{\dag}\bZ\ve{n}. \label{app12}
\eeq In destriping the amplitudes of the base functions are split
analogously: $\ve{a} = \ve{a}_p + \ve{a}_n$. The first component is
determined by the pixelisation noise and the second one by the
instrument noise.

Next we insert $\ve{y} = \ve{s} + \ve{n}$ into Eqs. (\ref{app5})
and (\ref{app6}). We obtain for the output map \bea
  \ve{m} = (\bP^T\bP)^{-1}\bP^T\ve{s} - (\bP^T\bP)^{-1}\bP^T\bH\widetilde{\ve{\Delta}}_p +\nonumber\\
  + (\bP^T\bP)^{-1}\bP^T[\ve{n} - \bH\widetilde{\ve{\Delta}}_n]. \label{app13}
\eea Ideally, we would like the output map of the map-making
algorithm to be equal to the first term on the right hand side. It
is called {\it binned noiseless map} in this study. The rest of the
terms bring error. They are represented by the {\it reconstruction
error map} (Tegmark~\cite{Teg97a}) \beq
   \beps = \ve{m} - (\bP^T\bP)^{-1}\bP^T\ve{s}. \label{app14}
\eeq The reconstruction error map is comprised of a signal component
($ \beps_p$) and a noise component ($ \beps_n$). \beq
   \beps =  \beps_p +  \beps_n.
  \label{app15}
\eeq \beq
  \beps_p = -(\bP^T\bP)^{-1}\bP^T\bH\widetilde{\ve{\Delta}}_p \label{app16}
\eeq
\beq
  \beps_n = (\bP^T\bP)^{-1}\bP^T[\ve{n} - \bH\widetilde{\ve{\Delta}}_n]
\label{app17} \eeq For destriping we can write analogously \beq
  \beps_p = -(\bP^T\bP)^{-1}\bP^T\bF\ve{a}_p \label{app18}
\eeq \beq  \beps_n = (\bP^T\bP)^{-1}\bP^T[\ve{n} - \bF\ve{a}_n].
\label{app19} \eeq

The signal and noise components of the reconstruction error map were
used extensively when the map-making algorithms were compared in
Sect.~\ref{sec:maps}. We studied the minimum, maximum and std values
of their pixel temperatures. Additionally, we produced their angular
power spectra and compared them as well.

\subsection{Map-making errors and pixelisation noise} \label{subsec:pixnoise}

The purpose of this section is to give a qualitative explanation to
the fact that the magnitude of the signal component of the
reconstruction error map is larger for the ML map-making than for
the destriping. The signal component of the reconstruction error
depends on $\widetilde{\ve{\Delta}}_p$ in the ML map-making and on
$\ve{a}_p$ in the destriping (see Eqs. (\ref{app16}) and
(\ref{app18})). The source of both quantities is the pixelisation
noise $\ve{p}$.

\begin{figure}[tbh]
\center{\resizebox{\hsize}{!}{\includegraphics{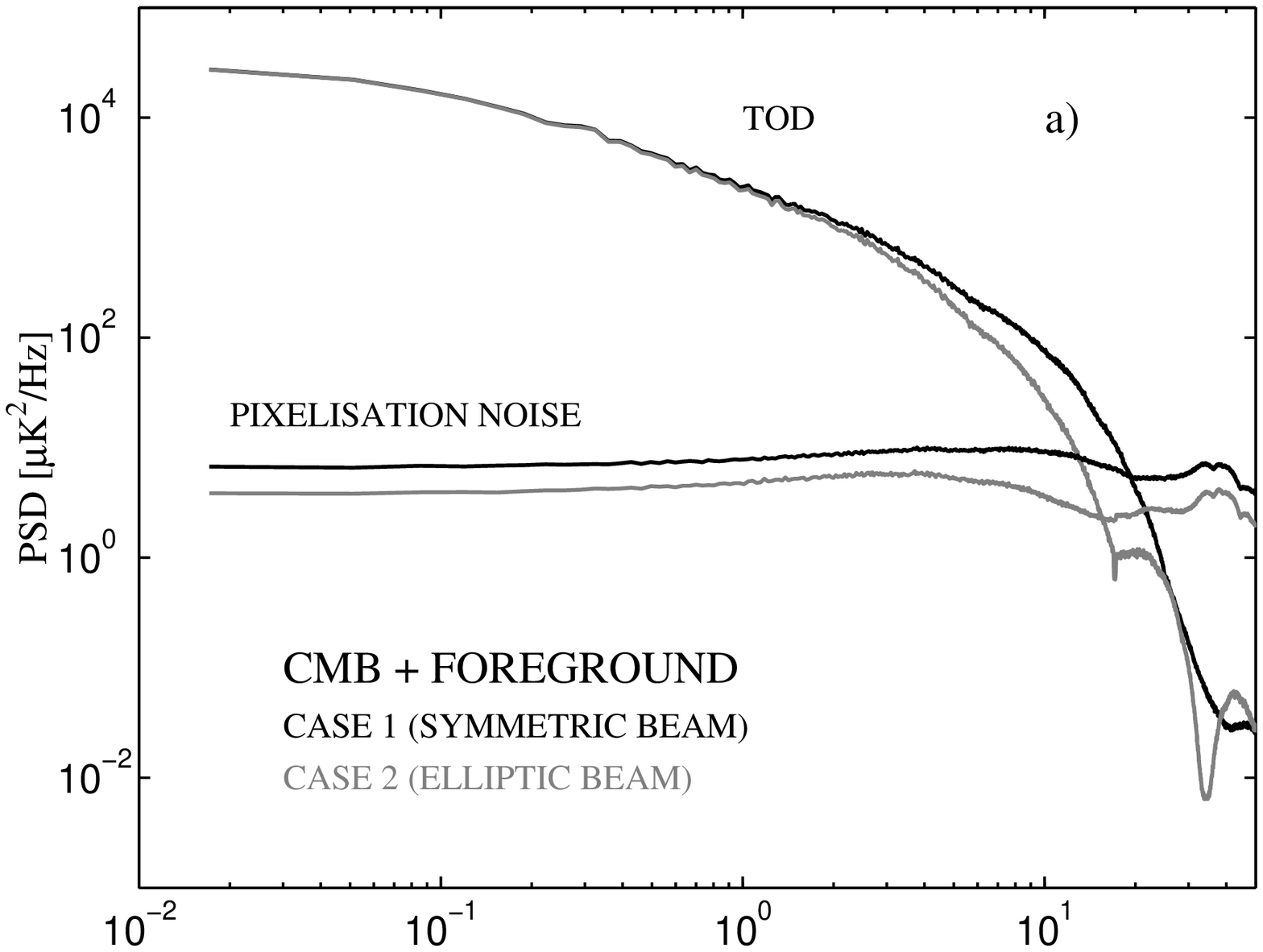}}}
\center{\resizebox{\hsize}{!}{\includegraphics{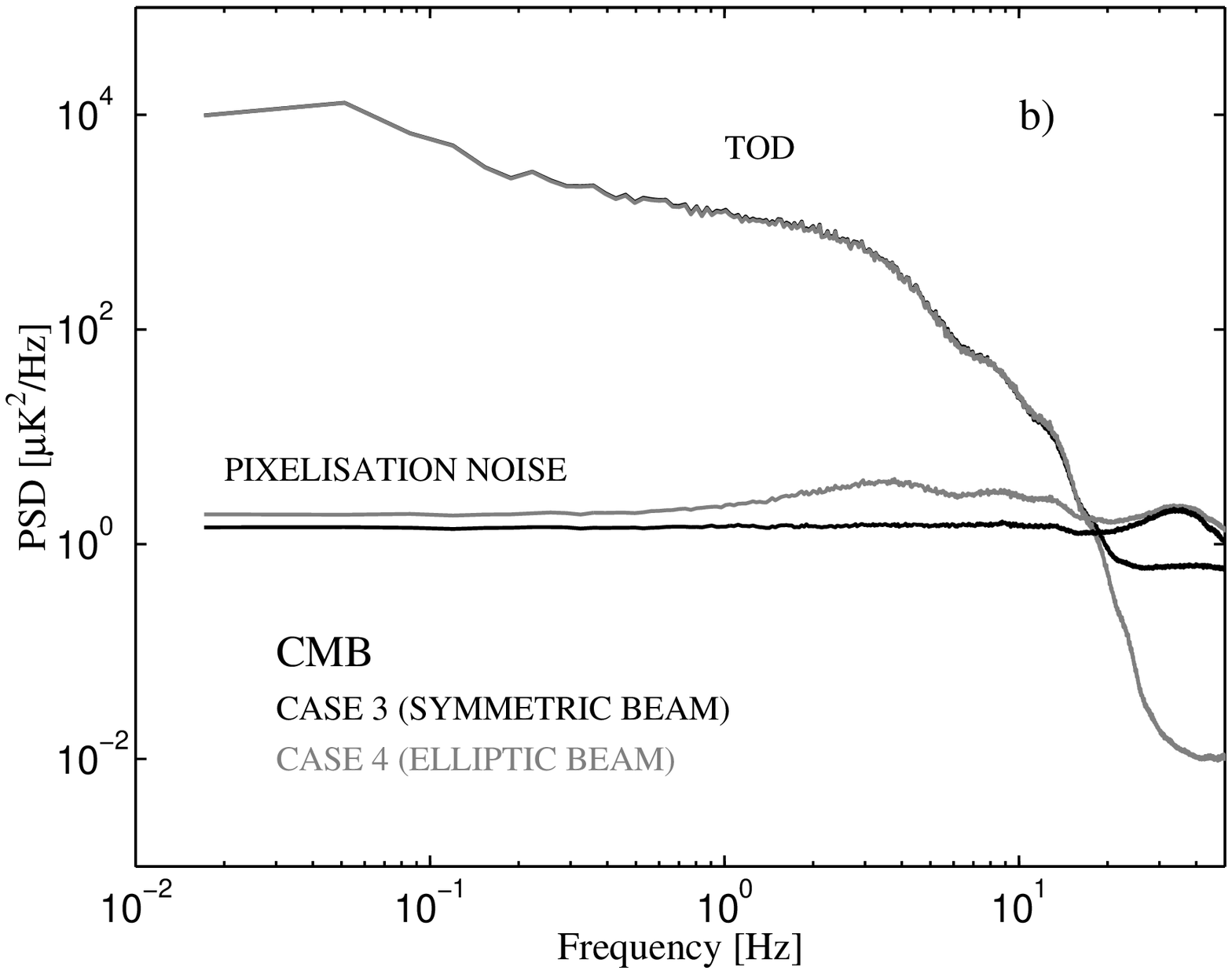}}}
\caption{The PSD of every 60 circle averaged ring of a simulated
signal-only TOD was calculated (8766 PSDs for a TOD) and the mean of
the PSDs was taken. The mean PSDs of all signal-only TODs generated
for this study are shown (curves labeled with "TOD"). The mean PSDs
of the associated pixelisation noise streams were calculated as well
and are shown too. {\bf a)} TOD contains CMB and foreground. {\bf
b)} TOD contains only CMB. The temperature units are antenna $\mu$K
at 100~GHz.} \label{pixnoise}
\end{figure}

We will first examine the spectrum of the pixelisation noise
($\tilde{\ve{p}} = \bH^{\dag}\ve{p}$). The PSD of every 60 circle
averaged ring of a simulated signal-only TOD was calculated and the
mean of these PSDs was taken over the full one year TOD. The mean
PSDs of all four simulated signal-only TODs and their associated
pixelisation noise streams ($\ve{p} = -\bZ\ve{s}$, $\ve{s}$ = TOD)
were determined. They are shown in Fig.~\ref{pixnoise}.

The level of the pixelisation noise is higher for the TODs
containing CMB and foreground than for the TODs of CMB only. This
explains why the signal component of the reconstruction error
increases when the foreground emissions are included in the
simulations.

The sample $p_k$ of the pixelisation noise stream ($k$ indexes the
sample) is
 \beq p_k = \left(\frac{1}{N_k}\sum_{i\in k}s_i\right) -
 s_k = \frac{1}{N_k}\sum_{i\in k}(s_i - s_k), \label{app23}
 \eeq
where $s_k$ is the k$^{\rm th}$ sample of the TOD, $i\in k$ refers
to those TOD samples (including $s_k$) that hit the same pixel as
$s_k$ and $N_k$ is the number of hits in that pixel. We can assume
that the magnitudes of the pixel-to-pixel correlations are smaller
for the temperature differences ($s_i - s_k$) than for the
temperatures themselves. This means that notable correlation exists
between $p_k$ and $p_{k'}$ only if they are samples from the same
pixel. Only the fraction $N_{\rm pix}/N_{\rm t} \ll 1$ of the TOD
samples are observations from the same pixel leading to, in average,
a weak correlation between the samples of the pixelisation noise.
This explains the "white noise" type nearly flat spectra of the
pixelisation noises (see Fig.~\ref{pixnoise}). The fact that in
Fig.~\ref{pixnoise}a the pixelisation noise is larger for the
symmetric beam than for the elliptic beam and opposite in
Fig.~\ref{pixnoise}b reflects the different methods that were used
in generating the symmetric beam TODs (total convolution in case 1
vs. scanning a high resolution map in case 3, see
Sect.~\ref{sec:tod}).

Vector $\widetilde{\ve{\Delta}}_p$ is a solution to Eq.
(\ref{app11}). The matrix $\widetilde{\bN}$ is diagonal with samples
(bins) of $1 + \fk/f$ (cf. Eq. (\ref{psd})) in its diagonals. The
diagonal elements of $(\widetilde{\bN} - \bIt)^{-1}$ are $\gg 1$ for
frequencies higher than the knee frequency. Fig.~\ref{pixnoise}
suggests that the power of
$\bH^{\dag}\bZ\bH\widetilde{\ve{\Delta}}_p$ is considerably smaller
than the power of $\widetilde{\ve{\Delta}}_p$ leading to an
approximation where we can ignore $\bH^{\dag}\bZ\bH$ in Eq.
(\ref{app11}). This indicates that in the ML map-making the signal
component of the reconstruction error is mainly determined by that
part of the pixelisation noise spectrum that falls below the knee
frequency (Hivon et al.~\cite{Hiv05}).

By looking at Eqs. (\ref{destr2}) and (\ref{app18}) we can expect
that in destriping the uniform baselines of the pixelisation noise
contribute to this error. Because we assumed an exact repetition of
the pointings of the 60 circles of a ring, the pixelisation noise of
those circles is periodic (with period $T$ = 60~s) and it has a
Fourier series representation whose Fourier mode frequencies are
multiples of $1/T$. The lowest (zero) mode contributes to the
uniform baselines, whereas the modes up to the knee frequency
(0.1~Hz) contribute to the reconstruction error of the ML
map-making. This explains why the signal component of the
reconstruction error is smaller in destriping than in ML map-making.
The evaluation of the exact effect in the maps is complicated by the
scanning strategy.

\section{Pixel window and pointing distribution effects}
\label{sec:inputmap}

There are a number of effects contributing to the difference between
the input spectrum $C_\ell^{\rm{in}}$ and that of the binned
noiseless map $C_\ell^{\rm{B}}$.  For cases 1, 2 and 4 these include
the beam shape, spectrum smoothing due to sample interpolation from
the totalconvolver temperature grid (see Sect.~\ref{sec:tod}) and
the effective elongation of the beam due to sample integration. The
sample integration was simulated in Level S by generating multiple
(fast) samples at a higher sampling rate and the final output sample
of the detector (at sampling rate $f_{\rm s}$) was an average (with
equal weights) of the fast samples.

In case 3 the interpolation and sample integration errors do not
occur, because the signal part of the TOD was made by picking the
temperatures (at sampling rate $f_{\rm s}$) from a high-resolution
{\em input map}, which was smoothed by a spherical beam.  Thus the
only beam effect was that of the spherical beam, which we already
corrected for in Fig.~\ref{clb}. The input spectrum displayed in
Fig.~\ref{clb} is actually that of this input map, corrected for the
beam.

\begin{figure}[tbh]
\center{\resizebox{\hsize}{!}{\includegraphics[width=7.5cm,height=6cm]{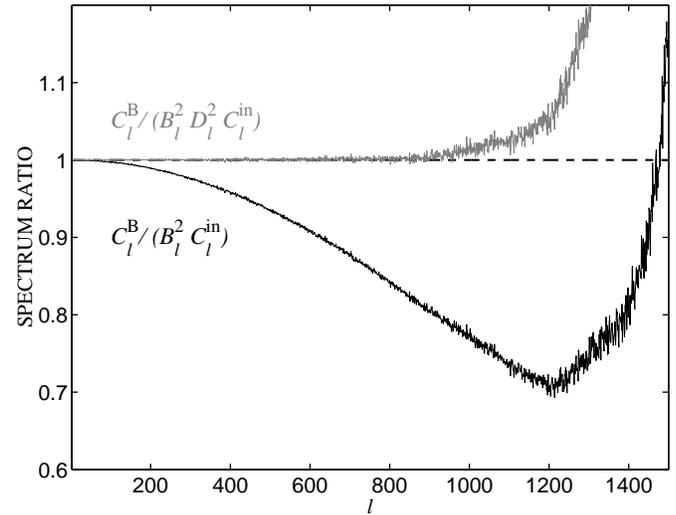}}}
\caption{The ratio of the spectra shown in Fig.~\ref{clb} (black
curve). The grey curve shows the same ratio after the spectrum of
the binned map ($C_\ell^{\rm B}$) has been deconvolved with an
approximate pixel window of the binning: $D_\ell^2 =
D_\ell^2(512)/D_\ell^2(1024)$.} \label{Bclb}
\end{figure}

Another effect comes from how the detector pointings sample the sky,
or, in this case (case 3), the small pixels of the input map, to
produce the binned map with its larger pixel size (the same as the
output map). In case 3 the input map had $N_\mathrm{side} = 1024$,
whereas the binned map had $N_\mathrm{side} = 512$, so that each
pixel of the binned map can be divided into four subpixels
corresponding to the pixels of the input map.  (Note that the
discussion in Sect.~\ref{sec:method} on map-making methods assumed
the same pixel size for input and output maps, and therefore did not
recognize the effects discussed here.) If each of the 4 subpixels
had been hit by the same number of times, the resulting binned
noiseless map would be just the input map downgraded to
$N_\mathrm{side} = 512$. The effect on the map spectrum should then
be given by the ratio of the two pixel windows
$D_\ell^2(512)/D_\ell^2(1024)$. Here $D_\ell(512)$ and
$D_\ell(1024)$ are the HEALPix pixel window functions for $N_{\rm
side}$ = 512 and $N_{\rm side}$ = 1024 (G\'orski et
al.~\cite{Gor05b}). (In the real situation the input map is replaced
by the sky with ``$N_\mathrm{side} = \infty$'', so that the
corresponding factor is just the pixel window of the binned map.) We
show in Fig.~\ref{Bclb} how this represents the effect well up to
$\ell \sim 800$.

The remaining effect, which blows up at high $\ell$, is due to two
things: 1) the nonuniform sampling of the four subpixels (or, in the
real world, that of the output map pixel area on the sky), 2) that
the HEALPix pixel window functions themselves represent an
approximation, as discussed below. This remaining effect represents
coupling between the $\ell$ modes of the spectra, which couples
power from the low-$\ell$ to the high-$\ell$ that shows up as an
high-$\ell$ excess power. This effect was discussed in Poutanen et
al.~\cite{Pou04}, where it was modelled as a signal bias.

Let us examine the distribution of the detector pointings in the sky
and its impact on the spectrum of the binned noiseless map in more
detail. The following discussion can be applied both to the real
case of observing the sky and the case (our case 3) where the TOD is
picked from a high-resolution pixelized input map. We consider the
samples $s_i$ ($i$ indexes the sample) of the CMB-only TOD that fall
in a pixel $k$ of the binned (or output) map (see Eq.
(\ref{app23})). The number of hits in that pixel is $N_k$. We assume
that every pixel is hit (100\% sky coverage at $N_{\rm side} = 512$
resolution), so that $N_k \ge 1$. The temperature of $s_i$ can be
given as
 \beq s_i = \sum_{\ell m}{a_{\ell m}B_\ell Y_{\ell m}(\ve{n}_i)}.
 \label{eqb_1} \eeq
Here $a_{\ell m}$ represent the CMB sky (see
Sect.~\ref{subsec:powerspectrum}), $B_\ell$ is the response of the
symmetric beam and $\ve{n}_i$ is a unit vector pointing in the
direction of the beam centre (or, in the case where the TOD is just
picked from an input map, the direction to the centre of the input
map pixel the detector is pointing at). The temperature of the pixel
$k$ of the binned map is
 \beq T_k^{\rm B} = \frac{1}{N_k}\sum_{i\in k}{s_i} =
 \sum_{\ell m} a_{\ell m} B_\ell\frac{1}{N_k}\sum_{i\in k}{Y_{\ell m}(\ve{n}_i)},
 \label{eqb_2} \eeq
where $i\in k$ refers to those TOD samples that hit the pixel $k$.

The expansion coefficients of the binned map are obtained by an
inverse spherical harmonic transformation
 \beq a_{\ell m}^{\rm B} = \Omega_p \sum_{k=0}^{N_{\rm pix}-1}{T_k^{\rm B} Y^{\ast}_{\ell m}(\ve{q}_k)}.
 \label{eqb_3} \eeq
We assume a HEALPix pixelisation, where the pixels have the same
area $\Omega_p = 4\pi/N_{\rm pix}$. The unit vector pointing to the
centre of the pixel $k$ is $\ve{q}_k$.

After inserting $T_k^{\rm B}$ from Eq. (\ref{eqb_2}) to Eq.
(\ref{eqb_3}) we obtain for the expansion coefficients of the binned
map
 \beq a_{\ell m}^{\rm B} = \sum_{\ell' m'}{a_{\ell' m'}B_\ell' \Omega_p \sum_{k=0}^{N_{\rm
 pix}-1}{\frac{1}{N_k} \sum_{i\in k} {Y_{\ell' m'}(\ve{n}_i) Y^{\ast}_{\ell m}(\ve{q}_k)}}}.
 \label{eqb_4} \eeq
This equation defines a coupling matrix
 \beq K_{\ell m \ell' m'}^{\rm B} \equiv \Omega_p \sum_{k=0}^{N_{\rm
 pix}-1}{\frac{1}{N_k} \sum_{i\in k} {Y_{\ell' m'}(\ve{n}_i) Y^{\ast}_{\ell m}(\ve{q}_k)}}
 \label{eqb_5} \eeq
between the $a_{\ell m}$ of the binned map and the CMB sky.

Using the statistical isotropy of the CMB sky ($\langle a_{\ell m}
a^{\ast}_{\ell' m'} \rangle = \delta_{\ell \ell'}\delta_{mm'}\langle
C_\ell^{\rm in} \rangle$) we obtain for the ensemble mean of the
angular spectrum of the binned map
 \beq \langle C_\ell^{\rm B} \rangle =
 \frac{1}{2\ell+1}\sum_{m=-\ell}^{\ell}{\langle |a_{\ell m}^{\rm
 B}|^2 \rangle} = \sum_{\ell'}{M_{\ell \ell'}^{\rm B}B_{\ell'}^2\langle C_{\ell'}^{\rm in} \rangle}, \label{eqb_6} \eeq
where $M_{\ell \ell'}^{\rm B}$ is the mode coupling matrix (kernel
matrix) of the binned map
 \beq M_{\ell \ell'}^{\rm B} = \frac{1}{2\ell+1}\sum_{m,m'=-\ell,-\ell'}^{\ell,\ell'}{|K_{\ell
 m \ell' m'}^{\rm B}|^2}. \label{eqb_7} \eeq

In spite of the fact that the binned noiseless map has a full sky
coverage, its mode coupling matrix $M_{\ell \ell'}^{\rm B}$ is not
diagonal but it is only diagonally dominant with small non-zero
off-diagonal elements, because the pixel area has been nonuniformly
sampled (in case 3, hits are in 4 subpixel centres only and unevenly
distributed among them). The off-diagonal elements are responsible
for the coupling of the power from the low-$\ell$ to high-$\ell$
that shows up as a high-$\ell$ excess power in $C_\ell^{\rm B}$ (see
Figs.~\ref{clb} and~\ref{Bclb}).

Finally, let us consider what happens if the number of hits in the
pixel increases, the hits in the pixel area become evenly
distributed and a symmetric circular pixel shape is assumed. In that
case the sum $\frac{1}{N_k} \sum_{i\in k} {Y_{\ell' m'}(\ve{n}_i)}$
can be approximated by an integral whose value can be given in a
simple form
 \beq
 \frac{1}{N_k} \sum_{i\in k} {Y_{\ell' m'}(\ve{n}_i)} \rightarrow
 D_{\ell'}Y_{\ell'm'}(\ve{q}_k), \label{eqb_8} \eeq where $D_{\ell'}
 \approx D_{\ell'}(512)$. (To be precise, the above limiting value will be reached, with
a different $D_\ell$, whenever the distribution of the hits is the
same in every observed pixel and the distribution is fully symmetric
around its centre $\ve{q}_k$). Under these assumptions we obtain an
approximation for the mode coupling matrix of the binned map
 \beq M_{\ell \ell'}^{\rm B} \approx D_{\ell'}^2M_{\ell \ell'}. \label{eqb_9} \eeq
Here $M_{\ell \ell'}$ is the mode coupling matrix of the MASTER
method (Hivon et al.~\cite{Hiv02} and
Sect.~\ref{subsec:powerspectrum} of this paper). We can see that the
MASTER approach for the power spectrum estimation (Eq. (\ref{aps1}))
corresponds to an approximation that a large number of hits is
symmetrically distributed in every observed pixel. For the full sky
map (like our binned noiseless map at $N_{\rm side}$ = 512
resolution) the MASTER mode coupling matrix $M_{\ell \ell'}$ is
close to a unit matrix and it cannot explain the high-$\ell$ excess
power that we see in the spectrum of the binned noiseless map. (The
full sky $M_{\ell \ell'}$ does have tiny non-zero off-diagonal
elements, because the spherical harmonics are not exactly an
orthogonal set of functions in the pixelised sky).

\clearpage

\end{document}